\documentclass[5p,number,sort&compress]{elsarticle}
 
\newcommand{\Lm}{\mathcal{L}_m}
\newcommand{\Hub}{\mathcal{H}}
\newcommand{\FF}{\mathcal{F}}
\newcommand{\dHS}{d_{\rm HS}}
\newcommand{\bHS}{b_{\rm HS}}

\usepackage{amssymb}
\usepackage{lipsum}
\usepackage{amsmath}
\usepackage{hyperref}
\usepackage{mathtools}
\usepackage{amsthm}
\usepackage{graphicx}
\usepackage{booktabs}
\usepackage{orcidlink}
\usepackage{cuted}

\hypersetup{
  colorlinks   = true, 
  urlcolor     = blue, 
  linkcolor    = blue, 
  citecolor   = blue 
}

\journal{Physics of the Dark Universe}
\begin{document}
\begin{frontmatter}

\title{\texorpdfstring{Implications of $f(R)$ gravity on late-time cosmic structure growth through a complete description of density perturbations}{Implications of f(R) gravity on late-time cosmic structure growth through a complete description of density perturbations}}

\author[first,second]{Miguel Barroso Varela \orcidlink{0009-0006-9844-7661}\texorpdfstring{\corref{cor1}}{cor1}}
\ead{up201907272@edu.fc.up.pt}

\author[third,fourth]{Álvaro de la Cruz-Dombriz \orcidlink{0000-0002-7072-9396}}
\ead{alvaro.dombriz@usal.es}
\affiliation[first]{organization={Departamento de Física e Astronomia, Faculdade de Ciências, Universidade do Porto},
            addressline={Rua do Campo Alegre s/n}, 
            city={Porto},
            postcode={4169-007}, 
            country={Portugal}}
\affiliation[second]{organization={Centro de Física das Universidades do Minho e do Porto},
            addressline={Rua do Campo Alegre s/n}, 
            city={Porto},
            postcode={4169-007}, 
            country={Portugal}}

\affiliation[third]{organization={Departamento de Física Fundamental, Universidad de Salamanca, 37008 Salamanca, Spain}}
\affiliation[fourth]{organization={Cosmology and Gravity Group, Department of Mathematics and Applied Mathematics, University of Cape Town, Rondebosch 7700, Cape Town, South Africa}}
            
\cortext[cor1]{Corresponding author.}

\begin{abstract}
We provide insight about the full form of the equations for matter density perturbations and the scalar Bardeen metric potentials in general $f(R)$ theories of gravity. 
When considering viable modifications to the standard $\Lambda$CDM background, the full scale-dependent equations for the metric perturbations are provided and are shown to match the ones obtained with the quasistatic approximation. We investigate the impact of the $n=2$ Hu-Sawicki model on the late-time growth of structures. We find that updated late-time growth of structure data imposes $|f_{R_0}|\lesssim10^{-6}-10^{-5}$ and thus conclude that the Hu-Sawicki $f(R)$ model contributes no significant phenomenology at both background and perturbative level beyond the effective cosmological constant encompassed in its definition. This conclusion points to the survival of the present tension between early and late measurements of $\sigma_8$, as the Hu-Sawicki model can only worsen this issue or at best reproduce the results from the current concordance cosmological model. The generalized perturbative method showcased in this work can be applied to more elaborate $f(R)$ models to isolate genuine higher-order signatures beyond the quasistatic approximation.
\end{abstract}




\end{frontmatter}




\section{Introduction}\label{sec:Introduction}

The past few years have seen the cosmology research community faced with the possible fragility of the Concordance cosmological model ($\Lambda$CDM). The perfect example of this paradigmatic shift is the so-called Hubble tension \cite{DiValentino:2021izs}, i.e., the seemingly ever-growing gap between early and late-time Hubble constant measurements, which has led to different ideas emerging in the literature, some aimed at critiquing the astrophysical aspects of direct observations, while others tackle more foundational aspects of either the underlying matter content or the gravitational theory. The possible detection of dynamical dark energy by the Dark Energy Spectroscopic Instrument (DESI) collaboration has only added fuel to the fire \cite{DESI:2024mwx,DESI:2025zgx}, sparking considerable debate on the origin of this as of yet unexplained behavior of what was thought to stem from a cosmological constant, which was problematic in its own right \cite{Weinberg:1988cp}.
In what regards modifications to the standard cosmological model, one can go one step further and separate them into background and perturbative effects, as some theories aim to directly alter the expansion history of the Universe via the homogeneous and isotropic Friedmann equation, while others attempt to tweak the evolution of the perturbations that give rise to many of the macroscopic effects we observe in the present, at the cost of typically much more complex inter-dependencies and nuanced effects in the underlying mathematical analysis. 
\par 
In the realm of alternatives to $\Lambda$CDM there are several promising avenues of research, such as the introduction of new matter content like the generalized Chaplygin gas \cite{Bilic:2001cg,Kamenshchik:2001cp,Bento:2002ps}, massive neutrinos \cite{Lesgourgues:2006nd}, axions \cite{Rogers:2023ezo} and quintessence scalar fields \cite{Zlatev:1998tr,Tsujikawa:2013fta}, amongst many others. Alternatively, modified theories of gravity are another possible source of new physics that may provide explanations for current tensions and puzzling observations. An emblematic example are $f(R)$ theories of gravity ({\it c.f.} \cite{DeFelice:2010aj} for extensive reviews and references therein), where the gravitational Lagrangian is assumed to take a non-trivial form of some function of the curvature scalar $R$. Due to their simplicity, these theories have been extensively researched in the context of cosmological expansion history (background) dynamics \cite{Hu:2007nk,delaCruz-Dombriz:2015tye}, their implications on black holes and other compact objects \cite{delaCruz-Dombriz:2009pzc,Cembranos:2012fd,Astashenok:2017dpo}, and inflationary mechanisms \cite{Starobinsky:1980te,Rinaldi:2014gua,Inagaki:2019hmm}. Other examples include more complex generalisations of this functional dependency to other quantities, as in $f(Q)$ \cite{Anagnostopoulos:2021ydo,Dimakis:2022rkd,Sahlu:2024dxp}, $f(T)$ \cite{Chen:2010va,Myrzakulov:2010vz,Cai:2015emx} and $f(R,\Lm)$ \cite{Bertolami:2007gv,BarrosoVarela:2024htf,BarrosoVarela:2024ozs} theories, along with other combinations of these, all with varying degrees of success and physical motivation.
\par
In the context of $f(R)$ theories, the choice of the function is the object of individual debate. Indeed, depending on the intended goal of the modification, there are several constraints that need to be obeyed for mathematical and physical consistency of the theory. After their early role in providing alternative mechanisms for sourcing the late-time accelerated expansion of the Universe with simply negative powers of the curvature \cite{Carroll:2003wy}, more consistent models were proposed \cite{Starobinsky:2007hu,Tsujikawa:2007xu,Cognola:2007zu,Appleby:2007vb}. The Hu-Sawicki $f(R)$ model \cite{Hu:2007nk}, which we analyze in this work, obeys the correct high and low-curvature limits consistent with the $\Lambda$CDM model, introducing the cosmological constant as a consequence of a modified gravitational action instead of imposing it as an ad hoc fluid in the theory, all while avoiding instabilities and obeying local gravity tests. In this work, after taking Ref. \cite{delaCruz-Dombriz:2008ium} as a departing point, we consider the full form of the linearized field equations for general $f(R)$ theories and analyze their implications on density perturbations and structure growth for the particular case of the Hu-Sawicki model, which is of specific interest to this analysis, as its behavior cannot be distinguished from $\Lambda$CDM at background level but may provide non-trivial effects at the linearized level. We then constrain the model's parameters using growth of structure data, namely the product of the growth rate $f_g$ and the $\sigma_8$ parameter, as well as individual measurements of each quantity. These isolated $f_g$ and $\sigma_8$ data were shown to be of significant interest to constraining parameters in modified theories, namely Horndeski theories with the speed of gravitational waves equal to that of light, as considered in Ref. \cite{Perenon:2019dpc}.
\par
Throughout this work, we consider a spatially flat, homogeneous and isotropic cosmological background described by the Friedmann-Robertson-Walker (FRW) metric in comoving coordinates
\begin{equation}\label{eq:BackgroundMetric}
    {\rm d}s^2=a^2(\eta)\left[-{\rm d}\eta^2+{\rm d}x^2+{\rm d}y^2+{\rm d}z^2\right]\,,
\end{equation}
where comoving time $\eta$ is related to cosmic time as ${\rm d}t=a(\eta){\rm d}\eta$. We denote derivatives with respect to $\eta$ with primes as $\partial_\eta X\equiv X'$ and use the comoving Hubble parameter $\Hub=a'/a$. Higher-order derivatives are written as $(\partial_\eta)^NX\equiv X^{(N)}$. When considering perturbations to the background metric \eqref{eq:BackgroundMetric}, we exclusively analyze the scalar sector of perturbations. Thus, we write the perturbed line element in the Newtonian gauge (also dubbed longitudinal) as
\begin{equation}\label{eq:PerturbedLineElement}
\begin{aligned}
    {\rm d}s^2=a^2(\eta)&\big[\left.-(1+2\Phi){\rm d}\eta^2
    +(1-2\Psi)\delta_{ij}{\rm d}x^i{\rm d}x^j\big]\right.\, ,
\end{aligned}
\end{equation}
although we stress that different sign conventions can be found in the literature. In this gauge the metric perturbations $\Phi$ and $\Psi$, usually referred to as Bardeen potentials, have a clear connection with the gravitational potential in the Newtonian limit, where in our sign convention $\Phi=\Psi$ in such a limit. \par

This communication is organized as follows. In Section \ref{sec:LCDMPerturbations} we briefly review the properties of density fluctuations in General Relativity (GR) for the $\Lambda$CDM model. Then, in Section \ref{sec:FRGravity} we describe the $f(R)$ action and field equations at both background and perturbative level in Subsections \ref{subsec:BackgroundData} and \ref{subsec:fR_Perturbations}, along with a discussion on the metric potentials, the validity of the quasistatic approximation against the full higher-order equations and their modifications of the matter power spectrum in Subsections \ref{subsec:MetricPotentials}, \ref{subsec:QuasistaticComparison} and \ref{subsec:PowerSpectrum} respectively. 
Subsequently, in Section \ref{sec:CosmologicalImplications} we discuss several cosmological implications of $f(R)$ gravity, including the velocity correlation functions and the integrated Sachs-Wolfe effect, among others. We conclude by fitting the Hu-Sawicki model to the latest growth of structure data catalog in Section \ref{sec:fsigma8Fit} and present our conclusions in Section \ref{sec:Conclusions}.  
We use the $(-,+,+,+)$ signature and set $c=1$, as well as $\kappa^2=8\pi G=M_P^{-2}=1$.

\section{\texorpdfstring{Density perturbations in $\Lambda$CDM}{Perturbations in LCDM}}\label{sec:LCDMPerturbations}
The well-known Einstein field equations for GR in a Universe with a cosmological constant $\Lambda$ are 
\begin{equation}
    R^\mu_{\;\nu}-\frac{1}{2}\delta^\mu_{\;\nu} R+\Lambda\delta^\mu_{\;\nu}=T^\mu_{\;\nu} \, ,
\end{equation}
where the matter content is described by the stress-energy tensor $T_{\mu\nu}$. When considering the FRW metric shown in Eq. (\ref{eq:BackgroundMetric}) along with a perfect fluid with density $\rho$ and pressure $p$, we obtain the standard Friedmann equation  $\Hub^2=a^2(\rho+\Lambda)/3$ along with the conservation equation
\begin{equation}\label{eq:ConservationEq}
    \rho'=-3\Hub(\rho+p)=-3\Hub(1+c_s^2)\rho\, ,
\end{equation}
where we have assumed an equation of state $p=c_s^2\rho$ ($c_s^2=1/3$ for radiation and $c_s^2=0$ for non-relativistic matter). By perturbing both the metric and the matter content we get  
\begin{equation}\label{eq:GRPerturbedFieldEqs}
\delta R^\mu_{\;\nu}-\frac{1}{2}\delta^\mu_{\;\nu}\delta R=
\delta T^{\mu}_{\;\nu} \, ,     
\end{equation}
where the perturbed Ricci scalar can be written explicitly in terms of the two scalar metric perturbations\footnote{From this point forward we work in Fourier space and thus all quantities should be written with a subscript, such as $\Phi_k$. Although we suppress this for simplicity, this should be kept in mind when analysing all equations with spatial derivatives.}
\begin{equation}\label{eq:deltaR}
\begin{aligned}
    \delta R=-\frac{2}{a^2}\left[3\Psi''+6(\mathcal{H}'+\mathcal{H}^2)\Phi +3\mathcal{H}(\Phi'+3\Psi')\right.\\
    \left.-k^2(\Phi-2\Psi)\right]\,,
\end{aligned}
\end{equation}
and the full scalar sector stress-energy tensor is given by
\begin{equation} \label{eq:StresEnergyTensor}
    \begin{aligned}
        & T^\eta_{\;\eta} =\bar T^\eta_{\;\eta} +\delta T^\eta_{\;\eta} =-(\rho+\delta \rho)=-(1+\delta)\rho\,, \\
        &T^i_{\; \eta}= \delta T^i_{\;\eta} =-(\rho+ p)\partial^i v=-\rho(1+c_s^2)\partial^i v\,,\\
        &T^i_{\;j}= \bar T^i_{\;j}+\delta T^i_{\;j}=(p+\delta p) \delta^i_{\;j}=c_s^2(\rho+\delta\rho)\delta^i_{\;j}\, ,
    \end{aligned}
\end{equation}
with $v$ denoting the potential for velocity perturbations, the relative density contrast defined as $\delta=\delta\rho/\rho$ and the perturbed fluid equation of state satisfying $\delta p=c_s^2\delta\rho$, as we assume adiabatic pressure perturbations with adiabatic sound speed $c_s$ \cite{dodelson2003}. 

\par
In general, equations are written in Fourier space, with $\mathbf{\nabla} X\rightarrow i\mathbf{k}X$ and $k\equiv | \mathbf{k}|$. When analysing the resulting equations, it is common practice to consider the sub-Hubble limit, where one takes $k\gg\Hub$ such that the fields vary on scales that are significantly smaller than the Hubble horizon. In the context of $\Lambda$CDM, the comoving Hubble parameter at present is approximately $\Hub_0\sim2\times10^{-4}\allowbreak \ \text{Mpc}^{-1}$ and therefore we can take the sub-Hubble approximation to hold for small redshifts ($z\lesssim5$), as considered in this work in the context of the late Universe, as long as we focus on scales with $k\gtrsim0.01\ \text{Mpc}^{-1}$, i.e., around 1-2 orders of magnitude greater than the comoving Hubble parameter. Nonetheless, in Section \ref{subsec:fR_Perturbations} we will derive the full equations for the evolution of density perturbations, which may be used to capture the correct behavior at all scales where linearity of perturbations may be assumed.  \par 
The non-diagonal $(ij)$ components of the linearized field equations \eqref{eq:GRPerturbedFieldEqs} impose the equality of the metric potentials $\Phi=\Psi$. Also, in the sub-Hubble limit, the $(\eta\eta)$ component gives a Poisson equation 
\begin{equation}\label{eq:GRPoisson}
    \nabla^2\Phi=\nabla^2\Psi=-\frac{a^2\rho}{2}\delta\rho\Rightarrow\Phi_{\text{WL}}\equiv\frac{\Phi+\Psi}{2}=\frac{a^2\rho}{2k^2}\delta\,, 
\end{equation}
where we have defined the weak lensing (WL) potential $\Phi_{\rm WL}$. Additionally, by perturbing the conservation equations $\nabla_\mu T^{\mu\nu}=0$ we obtain two additional constraints that are completely independent of the gravitational action provided that the coupling between matter and curvature is minimal as in GR
\begin{align}
    \frac{\delta'}{1+c_s^2}-3\Psi'-k^2v&=0\,,\label{eq:EtaConservationEq} \\
    \frac{c_s^2}{1+c_s^2}\delta+\Phi+\Hub(1-3c_s^2)v+v'&=0\label{eq:SpatialConservationEq} \,.
\end{align}
For the remainder of this work, we assume a late-time cosmology with matter content dominated by non-relativistic matter, as well as dust matter perturbations,
and thus take $c_s^2=0$. Consequently, when the Universe's matter content is dominated by dust, by combining the equations \eqref{eq:EtaConservationEq} and
\eqref{eq:SpatialConservationEq}, one finds that the dust matter 
perturbations satisfy
\begin{equation}\label{eq:CombinedConservationEq}
    \delta''+\Hub(\delta'-3\Psi')+k^2\Phi-3\Psi''=0\, ,
\end{equation}
which will prove useful in Section \ref{sec:FRGravity}. The general differential equation governing the evolution of dust matter density perturbations in $\Lambda$CDM is obtained by algebraically removing all dependence on the metric and velocity potentials at the cost of obtaining higher-order terms in $k$, yielding
\begin{equation}\label{eq:FullGRDensityDiffEq}
\begin{aligned}
    &\delta''+\Hub\frac{4k^4-12k^2a^2\rho-18a^4\rho^2}{4k^4-6k^2a^2\rho+4\Hub^2}\\
    &-\frac{4k^4-6k^2(3a^2\rho-2\Hub^2)+18a^2\rho(a^2\rho-3\Hub^2)}{4k^4-6k^2a^2\rho-36\Hub^2}\frac{a^2\rho\delta}{2}=0\, ,
\end{aligned}
\end{equation}
which in the sub-Hubble limit reduces to the simpler and more typically used equation
\begin{equation}\label{eq:GRDensityDiffEq}
    \delta''+\mathcal{H}\delta'-\frac{1}{2}a^2\rho\delta=0.
\end{equation}
Note that both are second-order differential equations and describe a scale-independent evolution for $\delta$. Once the density contrast is obtained, it is usual to describe the large-scale structure evolution in terms of the growth rate $f_g\equiv\frac{{\rm d}\ln\delta}{{\rm d}\ln a}=\Hub\frac{\delta'}{\delta}$. In the context of $\Lambda$CDM cosmology, the equation above can be solved to a good approximation by the parametrisation $f_g(z)=\left[\Omega_m(z)\right]^\gamma$, with $\Omega_m(z)=H_0^2\Omega_{m,0}(1+z)/\Hub^2(z)$ and where the growth index has been constrained to $\gamma\approx0.55$ for $\Lambda$CDM. It is also typical to separate the spatial and temporal dependencies of the density fluctuations in terms of the linear growth factor $D(z)$ as $\delta(z,\mathbf{k})=D(z)\delta_k(\mathbf{k})$. 

\section{\texorpdfstring{$f(R)$ gravity}{f(R) gravity}}\label{sec:FRGravity}
We now consider the class of modified gravity theories described by the total action
\begin{equation}
S=\int {\rm d}^4x \sqrt{-g}\left[\frac{1}{2}f(R)+\Lm\right] \, 
    \label{eq:FRAction}
\end{equation}
where $f(R)$ is an arbitrary function of the Ricci scalar, $g$ is the determinant of the metric and $\Lm$ is the Lagrangian density for the matter fields. This reduces to the GR action when $f(R)=R$ and a cosmological constant can be included in the gravitational sector by using $f(R)=R-2\Lambda$. By varying the action with respect to the metric we obtain the modified field equations
\begin{equation}
FG_{\mu\nu}=T_{\mu\nu}+\Delta_{\mu\nu}F+\frac{1}{2}g_{\mu\nu}(f_1-FR)\, ,
    \label{eq:FieldEquations}
\end{equation}
where we have defined $\Delta_{\mu\nu}\equiv\nabla_\mu\nabla_\nu-g_{\mu\nu}\Box$ and $F={\rm d}F/{\rm d}R$. Due to the minimal coupling of matter to gravity in these theories, the conservation equation is unaltered from the GR and is thus given in FRW spacetimes by Eq. (\ref{eq:ConservationEq}) for each physical fluid. \par
Although the calculations in this section are completely general for all $f(R)$ models, when analysing concrete cosmological implications of these classes of theories and comparing their predictions with growth of structure data we will need to specify a particular model. In general, a viable $f(R)$ model must be able to reproduce the late-time accelerated expansion of the Universe, while recovering the standard $\Lambda$CDM results in the early Universe in a way that ensures the correct matter and radiation dominated epochs along with the correct properties for the Cosmic Microwave Background  (CMB). The latter condition is satisfied as long as $F<1$. Additionally, satisfying $\frac{{\rm d}^2f(R)}{{\rm d}R^2}=\frac{{\rm d}F}{{\rm d}R}>0$ for high curvatures ensures the stability of the model, as Dolgov-Kawasaki instabilities do not arise in this case. Also, in order to ensure a positive and finite effective gravitational constant, given in these theories by $G_{\rm eff}\propto F^{-1}$, we must also impose $F>0$ for all values of the curvature. The combination of all of these conditions implies that $F(R)$ must be a monotonically growing function of $R$ with $0<F(R)<1$ for all curvatures.
\par
A model that satisfies all of the aforementioned conditions is the well-known Hu-Sawicki (HS) model \cite{Hu:2007nk}
\begin{equation}\label{eq:HuSawickiModel}
     f(R)=R-H_0^2\frac{b_{\rm HS}(R/H_0^2)^n}{1+d_{\rm HS}(R/H_0^2)^n}\, ,
\end{equation}
where $\bHS$ and $\dHS$ are free dimensionless parameters and $n$ is a positive integer. The Hubble constant $H_0$ is used as a standard cosmological scale to normalize the effects of the curvature $R$ and is kept constant at a fiducial value of $H_0=70 $ km/s/Mpc for all of the analysis in this work, as this will have no effect on the behavior of the theory apart from rescaling $\bHS$ and $\dHS$. It is easy to see that for late times (small $R$) we recover standard GR, i.e., $\lim_{R\rightarrow0}f(R)=R$. By requiring this function to mimic the same behavior of the $\Lambda$CDM model at early times (large $R$), we must impose that it reproduces the cosmological constant. This is equivalent to fixing 
\begin{equation}\begin{aligned}\label{eq:HS_Limit}
   \lim_{R\rightarrow\infty}f(R)=R-2\Lambda&\Rightarrow \frac{b_{\rm HS}H_0^2}{d_{\rm HS}}=2\Lambda\\
   &\Rightarrow\frac{b_{\rm HS}}{d_{\rm HS}}=\frac{2\Lambda}{H_0^2}=6\Omega_\Lambda\sim4.2 \, ,
\end{aligned} \end{equation}
such that we may write the HS model in the large $R$ or large $\dHS$ limits as 
\begin{equation}\begin{aligned}\label{eq:HS_WeakLimit}
    f(R)=R-2\Lambda\frac{1}{1+\frac{H_0^{2n}}{d_{\rm HS}R^{n}}}&\approx R-2\Lambda+\frac{b_{\rm HS}H_0^{2}}{d_{\rm HS}^2}\left(\frac{H_0^2}{R}\right)^n\\
    &=R-2\Lambda+\frac{6\Omega_\Lambda H_0^2}{d_{\rm HS}}\left(\frac{H_0^2}{R}\right)^n\, ,
\end{aligned} \end{equation}
where we have introduced the approximation of weak modifications to the $\Lambda$CDM limit ($f_{\Lambda\rm CDM}=R-2\Lambda$), which in this notation is associated with large values of curvature and/or $\dHS$, such that in this limit $|f_{\rm HS}-f_{\Lambda\rm CDM}|\ll1$. In this regime we can calculate the quantity
\begin{equation}
\begin{aligned}
    F\equiv\frac{{\rm d}f}{{\rm d}R}&\approx1-\frac{6n\Omega_\Lambda}{d_{\rm HS}}\left(\frac{H_0^2}{R}\right)^{n+1}\\
    &\sim1-\frac{4n}{d_{\rm HS}}\left[\frac{3}{(1+z)^3+12}\right]^{n+1} \,,
\end{aligned}
\end{equation}
where we have used $H_0^2\sim(\rho+\Lambda)/3$ and $R\approx\rho+4\Lambda$ for weak modifications of the late Universe with respect to GR and took $\Omega_m=0.3$ and $\Omega_\Lambda=0.7$ as rough estimates. 
Throughout this work, we choose to fix the exponent in the HS model to be $n=2$, as in $n=1$ the two parameters $\{b_{\rm HS},d_{\rm HS}\}$ trivially combine into one single parameter, whereas for $n>2$ we find the modified gravity effects are increasingly more suppressed at both the background and perturbative levels due to the increased exponentiation of the fraction in the equation above. This choice means that the present value of $F$ can be estimated for the weak HS model as 
\begin{equation}
    F(z=0)\approx1-\frac{1}{8\dHS} \, ,
\end{equation} 
such that the typically cited value of $f_{R_0}\equiv F(z=0)-1$ can be identified as $|f_{R_0}|\approx(8\dHS)^{-1}$.

\subsection{Constraints from background data}\label{subsec:BackgroundData}

Before we analyze any perturbative behavior in the HS model, we must consider its background dynamics. In order to determine if the relationship between $\bHS$ and $\dHS$ is compatible with an effective cosmological constant as in Eq. (\ref{eq:HS_Limit}), as well as verifying that the weak modification approximation given in Eq. (\ref{eq:HS_WeakLimit}) holds, one must start by comparing the modified background evolution given by the field Eqs. (\ref{eq:FieldEquations}) with observations. Of course, if both of these constraints are favoured by the data, we may take the background to behave identically to $\Lambda$CDM as a good approximation, which diminishes the computation power and time required for the calculations in the remainder of this work. 
Previous analyzes of the background evolution in the HS model were performed in Refs. \cite{delaCruz-Dombriz:2015tye,Busti:2015xqa}. However, these focused on the best datasets at the time, which in the following 10 years have significantly increased in quantity and quality of data \cite{Briday:2021rrm,Scolnic:2021amr,Carr:2021lcj,Rubin:2023jdq}, allowing for tighter constrains on the cosmological background behavior in the Hu-Sawicki model. We thus reassess these past investigations in light of the most complete supernova distance moduli data catalog available at present, as we will discuss in what follows.
\par
The accuracy of fitting background dynamics generally improves when utilising varied data samples with mostly small relative uncertainties, such that a particularly useful data source are distance moduli catalogs of type IA supernovae (SNIa). In this work we use the broadest catalog made available by the Dark Energy Survey (DES) collaboration\footnote{Data available at \url{https://github.com/des-science/DES-SN5YR}}, consisting of 1635 photometrically-classified SNIa in the redshift range $0.1<z<1.3$ and complemented by 194 low-redshift SNIa in the range $0.025<z<0.1$ \cite{DES:2024jxu}.
This dataset provides values of the distance moduli $\mu(z)$, which can be theoretically calculated for a general background $H(z)$ as 
\begin{equation}\label{eq:DistanceModulusDefinition}
    \mu(z)=5\log_{10}\left(\frac{d_L(z)}{1 \ \text{Mpc}}\right)+25 \, ,
\end{equation}
where
\begin{equation}
    d_L(z)=(1+z)c\int_0^z\frac{{\rm d}z'}{H(z')} 
\end{equation}
is the luminosity distance.
Although this data does not include calibration for the SNIa absolute magnitude $M_B$, unlike the Cepheid variable calibration in the Pantheon+ sample, and is thus unable to break the degeneracy between $M_B$ and $H_0$, these parameters are irrelevant for the analysis of the background dynamics of the Hu-Sawicki model, for which only the shape of the evolution $E(z)\equiv H(z)/H_0$ is relevant. We thus follow the same method as in Ref. \cite{DES:2024fdw} and marginalize over the combined parameter $\mathcal{M}=M_B+5\log_{10}(c/H_0)$ when determining the $\chi^2$ value for each set of parameters. This is done by defining the marginalized $\tilde\chi^2$ value as 
\begin{equation}
    \tilde\chi^2_{\rm SNIa}=\chi^2_{\rm SNIa}-\frac{B^2}{c}+\ln\left(\frac{C}{2\pi}\right)\, ,
\end{equation}
where 
\begin{equation}
B=\sum_i(\mathbf{C}^{-1}_{\text{stat+sys}}\Delta\vec{\mathbf{   D}})_i
\end{equation}
and
\begin{equation}
C=\sum_i\sum_j\left[\mathbf{C}^{-1}_{\text{stat+sys}}\right]_{ij}
\end{equation}
are defined as in their original presentation in Ref. \cite{Goliath:2001af}, with $\Delta{\vec{\mathbf{D}}}_i=\mu_i-\mu_{\rm th}(z_i)$ denoting the difference between data and theoretical predictions, while $\mathbf{C}_{\text{stat+sys}}$ denotes the full systematic and statistical covariance matrix. \par
We generate constraints on the modified theory using a Markov chain Monte Carlo (MCMC) sampler in the \textsc{cobaya} package \cite{COBAYA_paper} and plot the posteriors using the \textsc{GetDist} package \cite{Lewis:2019xzd}. The convergence of the chains was determined by the generalized version of the $(R-1)$ Gelman-Rubin statistic built into the package, for which we kept \textsc{cobaya}'s default criterion of $(R-1)=0.01$. Priors for all parameters were chosen to be uniform over adequate ranges to ensure no bias was introduced, with $\dHS$ and $\bHS$ being allowed to vary from $\mathcal{O}(1)$ to $\mathcal{O}(10^4)$, including regions of the parameter space with strong modifications of GR and others for which the HS model has no distinction from GR at background level. Note that we do not allow negative values for $\dHS$, as it was found in Ref. \cite{delaCruz-Dombriz:2015tye} that for $n=2$ one is constrained to $d_{\rm HS}>0$ due to the singular nature of solutions with $d_{\rm HS}<0$. Due to the positivity of the cosmological constant, one also required $b_{\rm HS}>0$, as the ratio of these two parameters is proportional to $\Lambda$, which obeys $\Lambda\geq0$ in an acceleratingly expanding Universe as our own. Due to the HS model recovering the $\Lambda$CDM action for large curvatures, we may take our initial conditions from $\Lambda$CDM at sufficiently high redshifts, which due to the curvature evolving with redshift as $R\sim(1+z)^3$ means one can safely assume $z_i=5$ as a starting point for the numerical evolution of the $f(R)$ background equations.\par
\begin{figure}[t!]
    \centering
    \includegraphics[width=0.8\linewidth]{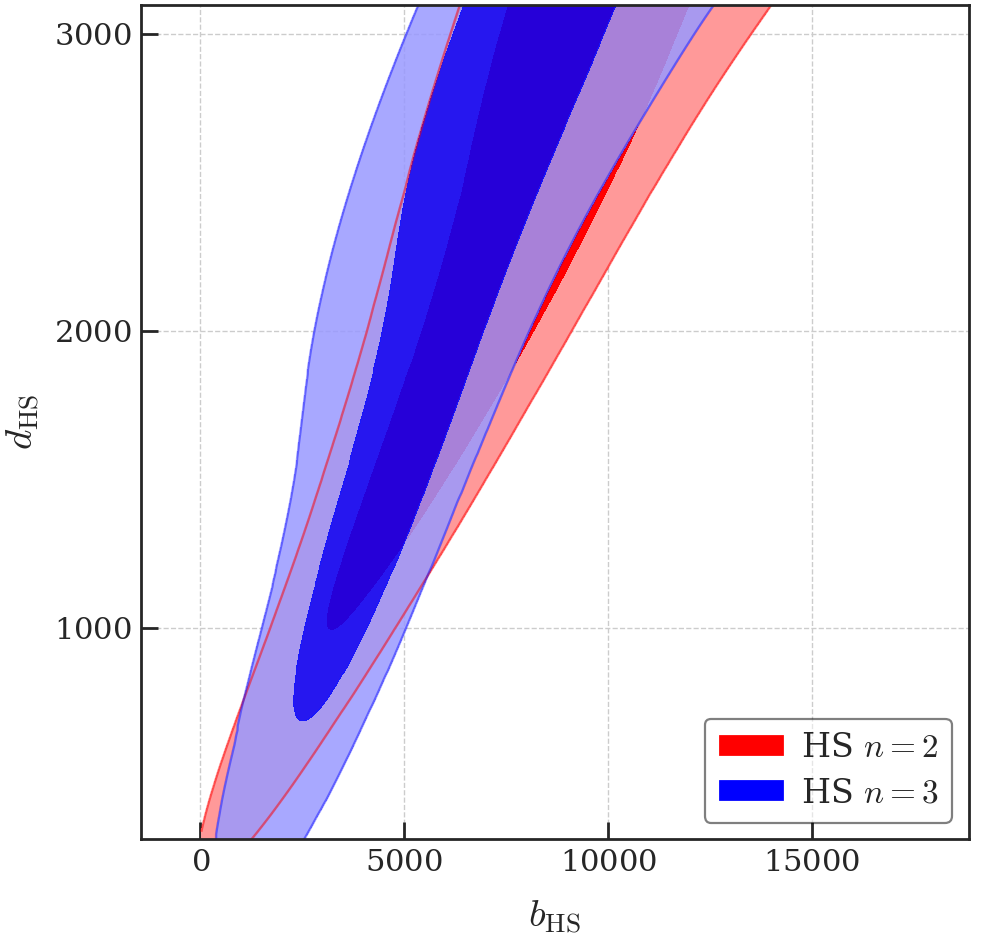}
    \caption{Posteriors for the parameters in the $n=2$ and $n=3$ Hu-Sawicki $f(R)$ models from fitting to SNIa distance moduli data from the DES collaboration. There is a clear linear correlation compatible with the presence of a cosmological constant in the modified gravitational action, with little effect from the choice of exponent $n$ in the model.}
    \label{fig:bHS_dHS_Correlation}
\end{figure}

As a consequence of all the above we are able to place a constraint on the correlation between the Hu-Sawicki parameters for $n=2$, as shown in Fig. \ref{fig:bHS_dHS_Correlation}. The weak convergence of the MCMC chains means that this ratio is not precisely fixed within the computing time and power available, but it is possible to place this value around $\bHS/\dHS\sim3-5\Rightarrow\Omega_\Lambda\sim0.5-0.8$, as expected for a $\Lambda$-dominated Universe in an accelerating expansion phase, in agreement with the standard result from the analysis of SNIa data when performed in pure $\Lambda$CDM cosmology.
We also find that within this correlated slice there is a preference for larger values of $\dHS$ and $\bHS$, pointing to weaker modifications of GR. However, we find the individual magnitudes of both $\bHS$ and $\dHS$ to be weakly constrained by background data, as for most values of reasonable magnitude ($\dHS,\bHS\gtrsim\mathcal{O}(100)$) one recovers a background evolution that is practically indistinguishable from $\Lambda$CDM. 
We have additionally analyzed the $n=3$ HS model, which shows a remarkable similarity in its parameter constraints to those found for $n=2$, showing that the aforementioned results are mostly independent of the exponent in the HS model, as expected due to the limit shown in Eq. \eqref{eq:HS_WeakLimit} having no dependence on $n$. 
This further motivates our choice of investigating uniquely the $n=2$ HS model, as previously discussed. It is worth noting that the weaker effects of the larger exponent are reflected in the weaker constraints on the magnitude of the HS free parameters, as seen in the corresponding posterior distribution. Considering all of this, we can be confident in fixing a constant ratio between $\bHS$ and $\dHS$ as in Eq. (\ref{eq:HS_Limit}) with $\Omega_\Lambda\approx1-\Omega_{m}$, such that the only free parameters in the $n=2$ HS model at both background and perturbative level for the late-time Universe are $\dHS$ and $\Omega_m$. 
We may thus also consider a background given by $\Lambda$CDM from large redshifts (large $R$) to the present, with its dynamics governed by the choice of $\Omega_m$ and $H_0$, although the precise value of the latter will be mostly irrelevant for the discussions in the rest of this work and will often be taken to have the fiducial value of $70$ km/s/Mpc. This will greatly simplify the complexity of calculations at the perturbative level, as all relevant modifications will come directly from the scalar first-order perturbed  $f(R)$ field equations, which we will discuss in what follows.

\subsection{\texorpdfstring{Density perturbations in $f(R)$ gravity}{Density perturbations in f(R) gravity}}\label{subsec:fR_Perturbations}
To analyze the perturbative dynamics in $f(R)$ gravity we must linearize the Field Eqs. (\ref{eq:FieldEquations})
\begin{equation}\label{eq:FRLinearisedFieldEq}
\begin{aligned}
 &F_R \delta R R_{\mu \nu}+F \delta R_{\mu \nu} -\frac{1}{2} g_{\mu \nu} F \delta R-\frac{1}{2} h_{\mu \nu} F\\
 &-\left[\delta(\nabla_\mu \nabla_\nu)-h_{\mu \nu} \square-g_{\mu\nu}\delta(\Box)\right]F\\
 &-\left[\nabla_\mu \nabla_\nu-g_{\mu \nu} \square\right]F_R \delta R = \delta T_{\mu \nu} \, ,
\end{aligned}
\end{equation}
where we have defined $F_R={\rm d}F/{\rm d}R$. Importantly, these are fourth-order differential equations, unlike the second-order equations obtained for GR in Section \ref{sec:LCDMPerturbations}. Analysing the off-diagonal $(ij)$ components shows that the metric potentials are no longer identical, instead differing by a quantity proportional to the curvature perturbations
\begin{equation}\label{eq:ijEq}
    \Psi-\Phi=\frac{F_{R}}{F}\delta R \, ,
\end{equation}
such that this equation now relates the quantities $\{\Phi,\allowbreak \Psi,\allowbreak \Psi',\allowbreak \Phi',\allowbreak \Psi''\}$, as per the $\delta R$ definition \eqref{eq:deltaR} in terms of the Bardeen potentials and their derivatives. On the other hand, the $(\eta\eta)$ and $(i\eta)$ components, respectively, give the following:
\begin{gather}
F\left[\Psi'+\Phi'+\Hub(\Psi+\Phi)\right]+F'(2\Phi-\Psi)=-\rho(1+c_s^2)v\,,\label{eq:iEtaEq} 
\end{gather}
\begin{equation}\begin{aligned}
    F\left[k^2(\Phi+\Psi)+3\Hub(\Phi'+\Psi')+(6\Hub^2-3\Hub')\Phi+3\Hub'\Psi\right]\\
    +3F'(\Psi'-\Hub\Psi-3\Hub\Phi)=-a^2\rho\delta\label{EtaEtaEq} \, .
\end{aligned} \end{equation}

Note that we do not show the diagonal $(ij)$ component of the linearized field equations, as the combination of both conservation equations, \eqref{eq:EtaConservationEq} and \eqref {eq:SpatialConservationEq},
with the above components can be used to obtain these, thus rendering them trivial, in agreement with the number of perturbative degrees of freedom. \par
Due to the higher-order nature of these equations, the resulting general differential equation for $\delta$, analogous to Eq. (\ref{eq:FullGRDensityDiffEq}) in GR, will be a fourth-order differential equation. As introduced and detailed in Ref. \cite{delaCruz-Dombriz:2008ium}, by algebraically removing the $v$-dependence of the equations one can find lengthy expressions for both $\Phi(\delta,\delta',\delta'',\delta''')$ and $\Phi'(\delta,\delta',\delta'',\delta''')$, along with equivalent expressions for $\Psi$. By requiring the derivative of $\Phi$ to be equal to $\Phi'$ one can write 
\begin{equation}
\frac{{\rm d} \Phi(\delta,\delta'',\delta'',\delta''')}{{\rm d}\eta}=\Phi'(\delta,\delta',\delta'',\delta''') \,,
\end{equation}
which gives the required general fourth-order differential equation for $\delta$
\begin{equation}\label{eq:FullDiffEq}
\beta_4\delta^{(4)}+\beta_3\delta'''+(\alpha_2+\beta_2)\delta''+(\alpha_1+\beta_1)\delta'+(\alpha_0+\beta_0)\delta=0\,.
\end{equation}
Here we have used the same notation as in the original result from Ref. \cite{delaCruz-Dombriz:2008ium}, with $\alpha$ being the coefficients representing the GR terms also present in Eq. (\ref{eq:FullGRDensityDiffEq}) whereas the coefficients $\beta$ denote terms that are unique to modifications to GR and, therefore, vanish in the limit $F(R)=1$. When presenting these results, we adopt dimensionless variables $\kappa_i=\Hub^{(i)}/\Hub^{i+1}$ and $\FF_i=F^{(i)}/(\Hub^iF)$ in order to simplify the expressions. Note that $\FF_i$ is defined differently from the original results from Ref. \cite{delaCruz-Dombriz:2008ium}, where $f_R=F-1$ was used instead of $F$ in the denominator. 
\par
Once \eqref{eq:FullDiffEq} is at hand, under the sub-Hubble approximation ($k\gg\Hub$) we keep only the highest-order terms in $k$ for both $\alpha$ and $\beta$ coefficients and obtain the leading-order differential equation 
\begin{equation}\label{eq:LeadingDiffEq}
\begin{aligned}
    \delta''&+\Hub\delta'\\
    &+\frac{F^5\Hub^2(\kappa_1-1)(2\kappa_1-\kappa_2)-\frac{16}{a^8}F_R^4(\kappa_2-2)k^8a^2\rho}{F^5(\kappa_1-1)+\frac{24}{a^8}F_R^4F(\kappa_2-2)k^8}\delta=0\,,
\end{aligned}
\end{equation}
where $\beta_{3,4}$ terms are lower order in $k$ and thus are not present, reducing this to a second-order differential equation as its GR counterpart. When considering the GR limit, i.e., we set $F=1\Rightarrow F_R=0$, and use the results for a dust-dominated  GR background $\kappa_1=-1/2$ and $\kappa_2=1/2$, then Eq. (\ref{eq:GRDensityDiffEq}) is recovered. As in both cases the coefficients for $\delta''$ and $\delta'$ are identical up to a factor of $\Hub$, this equation can be thought of as analogous to its GR counterpart with an effective gravitational constant $G_{\rm eff}(z,k)$ that is both time and scale-dependent \cite{Tsujikawa:2007gd}.

\subsection{Metric potentials}\label{subsec:MetricPotentials}
As discussed above, we can write both $\Psi$ and $\Phi$ in terms of $\delta$ and its derivatives of up to third-order. Similarly to what was done for $\delta$, we can write these as 
\begin{align}
    \Phi&=\sum_{i=0}^3\mathcal{C}_{\Phi,i}(k,\kappa_j,\FF_j)\delta^{(i)}\,,\label{eq:PhiSum}\\
    \Psi&=\sum_{i=0}^3\mathcal{C}_{\Psi,i}(k,\kappa_j,\FF_j)\delta^{(i)}\label{eq:PsiSum}\,,
\end{align}
where the coefficients can no longer be fully separated in unique GR and $f(R)$ parts as they are fractions with denominators with contributions from both pure GR and pure non-GR effects. These coefficients are given in their abridged form in Appendix \ref{sec:CoefficientsAppendix}. Therein we introduce an additional dimensionless variable $\epsilon=\Hub/k$, which serves as a way to organize each coefficient into a power series of diminishing magnitude due to the smallness of $\epsilon$ in the sub-Hubble regime ($k\gg\Hub\Rightarrow\epsilon\ll1$). As expected, the coefficients reduce to the GR result when setting $\FF_n\rightarrow0$ and further simplify in the sub-Hubble limit, yielding the standard relation of the Poisson Eq. (\ref{eq:GRPoisson}). The dependence in \eqref{eq:PhiSum} and \eqref{eq:PsiSum} on third-order derivatives of $\delta$ is fully due to the higher-order nature of $f(R)$ theories, such that these coefficients necessarily vanish in the GR limit, while also being of lower-order in $k$ (and thus higher-order in $\epsilon$) than the remaining $f(R)$ coefficients. Although in what follows we will see how these compare to the corresponding quasistatic predictions, it is important to note that for general $f(R)$ theories these full equations can lead to non-trivial effects on both the density perturbations via Eq. (\ref{eq:FullDiffEq}) \cite{delaCruz-Dombriz:2008ium,delaCruz-Dombriz:2009cop} and the metric perturbations via Eqs. (\ref{eq:PhiSum}) and (\ref{eq:PsiSum}).

\subsection{Comparison with quasistatic approximation}\label{subsec:QuasistaticComparison}
So far all results have only assumed the sub-Hubble limit at most, which as we have seen is a good approximation for most scenarios of late-time cosmological interest. However, another common assumption in the literature is the quasistatic approximation, which assumes that the background and the metric potentials vary considerably slower than the density perturbations, such that one may neglect the effects of terms like $\Psi'$ in comparison to $\delta'$. As seen in Section \ref{sec:LCDMPerturbations}, in GR the sub-Hubble regime is equivalent to the quasistatic regime, as both lead to the well-known Eq. (\ref{eq:GRDensityDiffEq}) instead of Eq. (\ref{eq:FullGRDensityDiffEq}). The same cannot be said in $f(R)$ gravity, as the sub-Hubble limit for general modifications leads to Eq. (\ref{eq:LeadingDiffEq}) instead of the more commonly seen quasistatic result \cite{Zhang:2005vt,Tsujikawa:2007gd}
\begin{equation}\label{eq:QuasisaticDiffEq}
    \delta''+\Hub\delta'-\frac{F+4\frac{k^2}{a^2}F_R}{F+3\frac{k^2}{a^2}F_R}\frac{a^2\rho}{2F}\delta=0 \,.
\end{equation}
The modified terms in this equation are of lower order in $k$ and are thus suppressed in the sub-Hubble regime for general $f(R)$ theories. However, as shown in Ref. \cite{delaCruz-Dombriz:2008ium}, assuming weak modifications to GR (as defined in Section \ref{sec:FRGravity}) leads to the coefficients in Eq. (\ref{eq:FullDiffEq}) being reorganized into precisely the quasistatic equation above provided $k\gg\mathcal{H}$. The same happens to the equations for the metric perturbations, which in the weak $f(R)$ limit rearrange into the typical quasistatic expressions \cite{Zhang:2005vt,Tsujikawa:2007gd}
\begin{align}
    \Phi_{\rm QS}&=-\frac{F+4\frac{k^2}{a^2}F_R}{F+3\frac{k^2}{a^2}F_R}\frac{a^2\rho}{2F}\frac{\delta}{k^2}\,,\label{eq:PhiQS}\\
    \Psi_{\rm QS}&=-\frac{F+2\frac{k^2}{a^2}F_R}{F+3\frac{k^2}{a^2}F_R}\frac{a^2\rho}{2F}\frac{\delta}{k^2}\label{eq:PsiQS} \, .
\end{align}
Note that these expressions may be rewritten in terms of the modified gravity parameters $\Sigma$, $Q$ and $\eta$ defined\footnote{Note that in Ref. \cite{Amendola:2016saw} the metric perturbations $\Phi$ and $\Psi$ are switched around in comparison to the definitions in this work.} in Ref. \cite{Amendola:2016saw}. In terms of these definitions we may write these quantities in the quasistatic approximation as
\begin{align}
    Q(z,k)&\equiv-\frac{2k^2 \Psi}{a^2\rho\delta}=\frac{1+2\frac{k^2}{a^2}\frac{F_R}{F}}{F+3\frac{k^2}{a^2}F_R}\,,\\
    \eta(z,k)&\equiv\frac{\Psi}{\Phi}=\frac{F+2\frac{k^2}{a^2}F_R}{F+4\frac{k^2}{a^2}F_R}\,,\\
    \Sigma(z,k)&\equiv-\frac{k^2\Phi_{\rm WL}}{a^2\rho\delta}=\frac{1}{2}Q(1+1/\eta)=\frac{1}{F}\, ,
\end{align}
where $\Phi_{\rm WL}\equiv(\Phi+\Psi)/2$.
In fact, only two of these are required to sufficiently describe the effect of the modifications to GR on the metric perturbations, as the third may be calculated from the others. Although we do not plot these quantities in what follows, their values can be straightforwardly obtained from the above equations given a set of HS parameters and initial conditions for the cosmological background, as they are defined in terms of zeroth-order quantities.
\par
\begin{figure*}[ht!]
  \centering 
\includegraphics[width=0.405\linewidth]{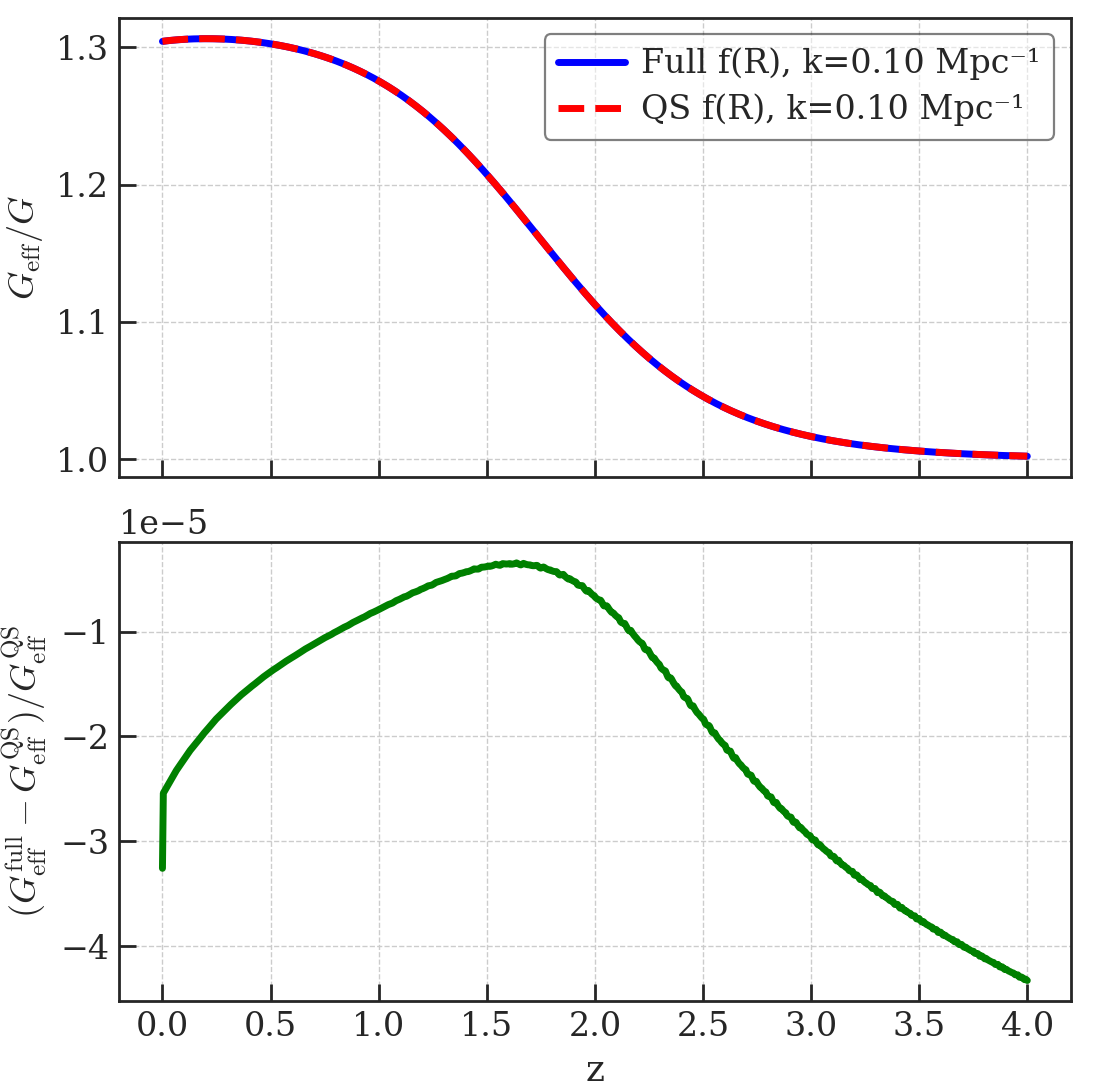}
\includegraphics[width=0.486\linewidth]{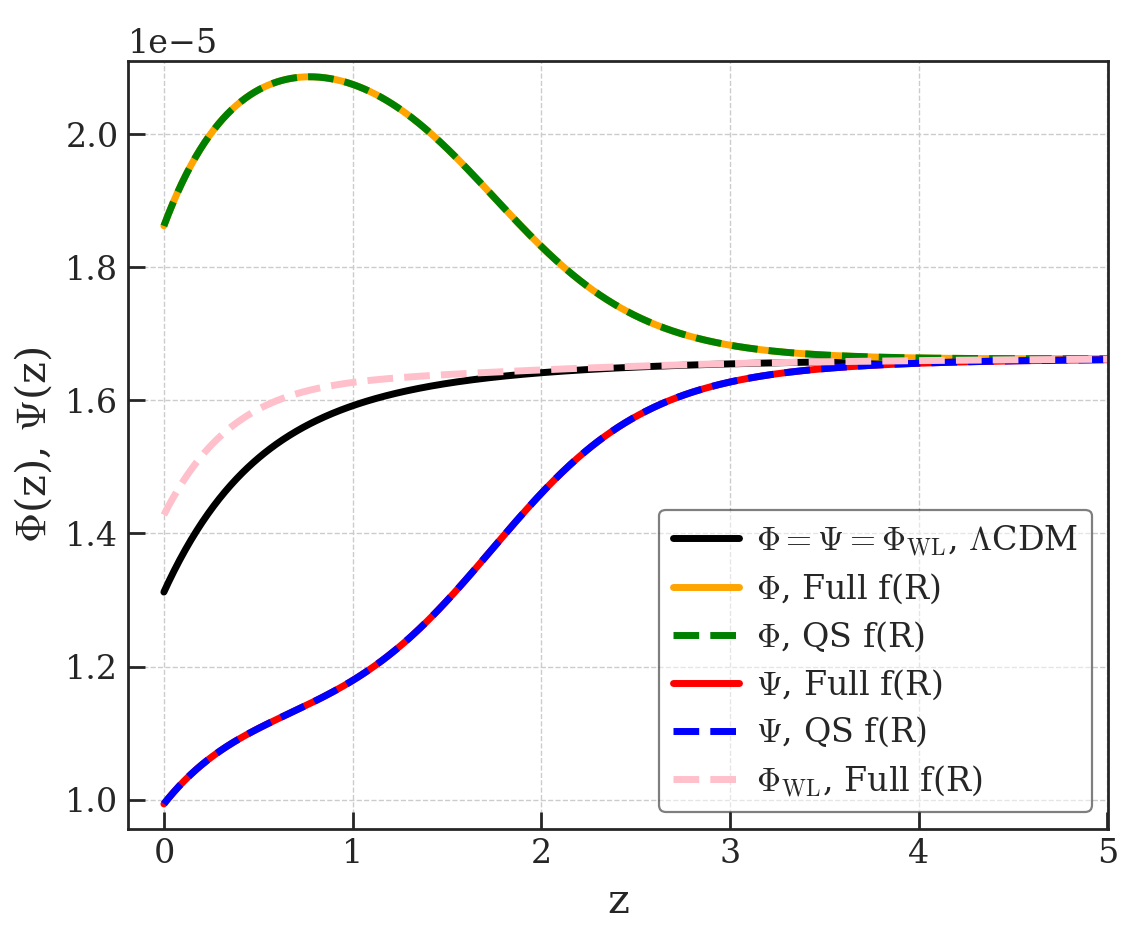}
    \caption{Comparison of the quasistatic and full equations for both the effective gravitational constant in the differential equation for $\delta$ (left) and the metric potentials (right). We consider the reference values of $k=0.1 \ \text{Mpc}^{-1}$ and $d_{\rm HS}=200$. The metric perturbations $\Phi$ and $\Psi$ differ from each other due to the non-linear nature of the gravitational action, as shown in Eq. \eqref{eq:ijEq}.
    } 
\label{fig:GEff_MetricPotentials}
\end{figure*}

Using the HS model defined in Eq. (\ref{eq:HuSawickiModel}) we may compare the full and quasistatic equations for different values of $\dHS$ and thus determine for which values the quasistatic approximation holds acceptably and for which ones we should use the full expression in Eq. (\ref{eq:FullDiffEq}). As seen in Section \ref{subsec:BackgroundData}, cosmic expansion history data shows a preference for values of $\dHS$ above $\sim10^2$. We therefore limit our analysis to this lower bound on $\dHS$, as even if lower values raised differences in the quasistatic and full expressions they would still be ruled out observationally at background level. We show the results for the effective gravitational constant for $d_{\rm HS}=200$ and $k=0.1 \ \text{Mpc}^{-1}$ in the left panel of Fig. \ref{fig:GEff_MetricPotentials}. As seen in the bottom left panel, the two approaches yield identical values for $G_{\rm eff}$ with relative differences of at most $\mathcal{O}(10^{-5})$, such that there is no practical distinction between the two for observational purposes, as we shall discuss in more detail in Section \ref{sec:CosmologicalImplications}. \par

In the right side of Fig. \ref{fig:GEff_MetricPotentials} we show results for both metric potentials $\Phi$ and $\Psi$, as well as the weak lensing potential $\Phi_{\rm WL}$, for both the quasistatic and full expressions. Although there is a clear deviation from GR in the individual potentials, the weak lensing potential has a much less pronounced difference, as the average of Eqs. (\ref{eq:PhiQS}) and (\ref{eq:PsiQS}) cancels out the leading higher-order effects in the sub-Hubble limit, being rescaled by $1/F$ and thus only slightly altered from the $\Lambda$CDM result, which for the set of parameters chosen for the example shown in the right side of Fig. \ref{fig:GEff_MetricPotentials} corresponds to a relative difference of around $+10\%$ at present. The matching of the quasistatic and full equations for the metric potentials is a natural extension of the same result for the density fluctuations presented in Ref. \cite{delaCruz-Dombriz:2008ium}, which is now confirmed under the same conditions, i.e., for weakly modifying $f(R)$ theories of gravity ($\vert F -1\vert \ll1$) the quasistatic, i.e. that from Eq. \eqref{eq:QuasisaticDiffEq}, and full higher-order behaviors at the perturbative level are indistinguishable provided $k\gg\mathcal{H}$.

\subsection{Modifications to the power spectrum}\label{subsec:PowerSpectrum}
The matter power spectrum as predicted by $\Lambda$CDM is calculated from the two-point correlation function of density fluctuations and thus evolves quadratically with the linear growth factor $D_{\Lambda{\rm CDM}}(z)$ from some initial redshift $z_i$ as 
\begin{equation}
    P_{\Lambda \rm CDM}(z,k)=P_{\Lambda \rm CDM}(z_i,k)\left(\frac{D_{\Lambda\text{CDM}}(z)}{D_{\Lambda\text{CDM}}(z_i)}\right)^2\, , 
\end{equation}
where $P(z_i,k)$ is the linear matter power spectrum at redshift $z_i$ computed from the primordial curvature spectrum. This quadratic dependence follows from the power spectrum being calculated from the variance of the Fourier modes of the matter density fluctuations, which individually scale as $D(z)$ by definition. However, when considering a modified theory of gravity with a scale-dependent growth factor $D(z,k)$, this rescaling modifies both the redshift and scale-dependence of the power spectrum. If the transition between the $\Lambda$CDM and the modified regimes occurs at some redshift $z_t$, i.e. $\Lambda$CDM represents the cosmological evolution well for $z>z_t$, we can interpolate between that point and smaller redshifts by using
\begin{equation}
    P_{{\rm HS}}(z,k)=P_{\Lambda \rm CDM}(z_t,k)\left(\frac{D_{\rm HS}(z,k)}{D_{\Lambda\text{CDM}}(z_t)}\right)^2 \, ,
\end{equation}
where for the HS model at perturbative level the transition redshift is approximately $z_t\lesssim5$ depending on the value of $\dHS$.

\begin{figure*}[ht!]
    \centering
    \includegraphics[width=0.8\linewidth]{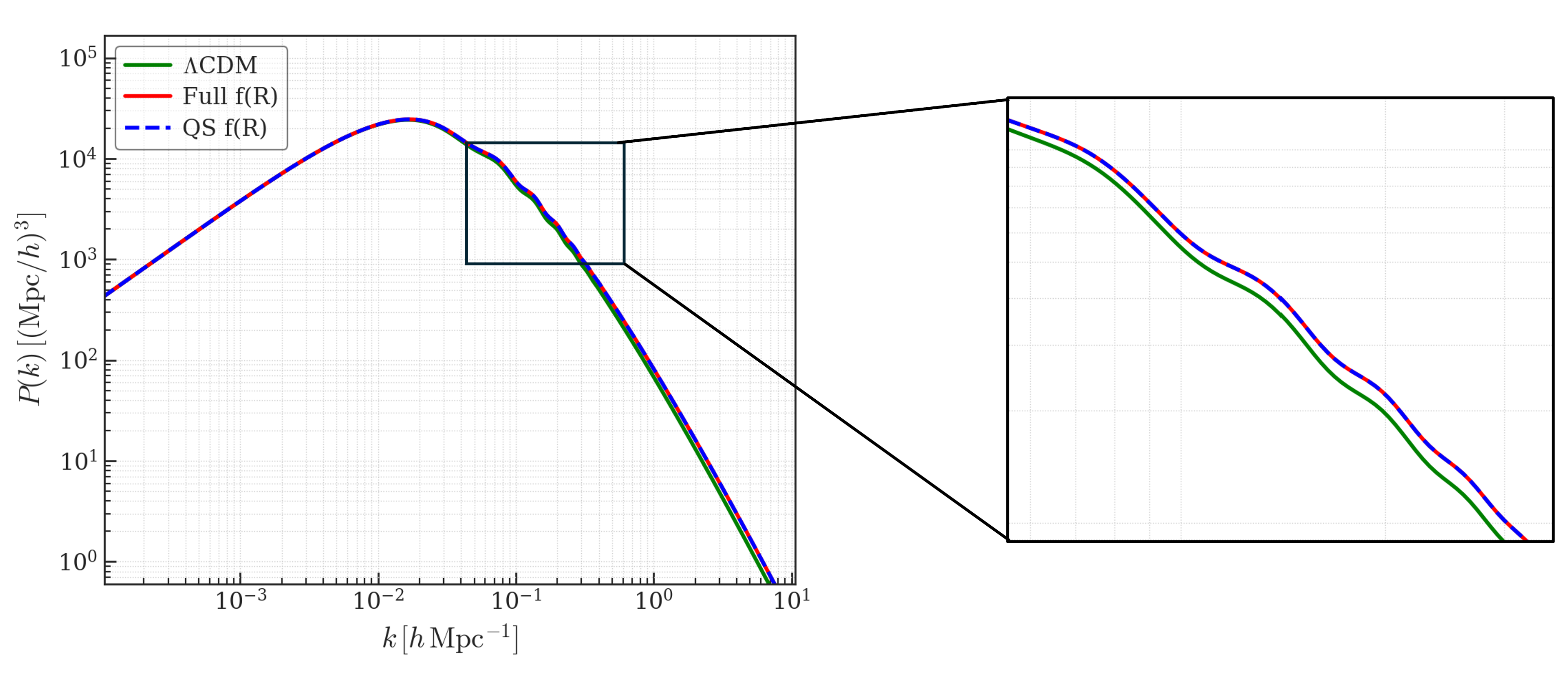}
    \caption{The modified matter power spectrum in the Hu-Sawicki $f(R)$ model for $z=0$. Fiducial values were used for the cosmological parameters, namely $\Omega_m=0.3$ and $H_0=70 \text{ km/s/Mpc}$. The quasistatic and full equations lead to indistinguishable  predictions.}
    \label{fig:PowerSpectrum}
\end{figure*}
We use the publicly available \textsc{camb} package \cite{Lewis:2002ah} to generate the power spectrum for fiducial cosmological parameters and modify it using the equation above. The results are shown Fig. \ref{fig:PowerSpectrum}. As expected, there are no significant deviations for $k\lesssim10^{-2}\ \text{Mpc}^{-1}$, with slight differences arising for smaller scales. These modifications have no tangible impact on most cosmological aspects, although in combination with other effects they can lead to distinct behavior from $\Lambda$CDM, as we shall see in Section \ref{sec:CosmologicalImplications}. Not surprisingly, the quasistatic approximation holds for the HS model in this scenario, as we don't consider values that considerably impact the background evolution of the Universe.

\section{Cosmological implications} \label{sec:CosmologicalImplications}
Our discussion thus far has focused on mostly theoretical aspects of the linearized $f(R)$ field equations. In this section, we analyze other concrete observational predictions that may be compared to present or future observational surveys of the growth of large-scale structures in order to constrain $f(R)$ theories that, while mimicking the $\Lambda$CDM background behavior, introduce non-trivial effects at the perturbative level.
In what follows, we consider the reference value for a HS model with moderately strong modifications at the perturbative level with $d_{\rm HS}=200$, associated with the (considerably high) value of $|f_{R_0}|\sim10^{-4}$ as defined in Section \ref{sec:FRGravity}, in order to highlight the effects of the modified theory, although as we shall see in Section \ref{sec:fsigma8Fit} this relatively small value of $\dHS$ (large $|f_{R_0}|$) is disfavoured by data. 

\subsection{Redshift-space distortion}
The connection between the redshift-space galaxy power spectrum and the power spectrum in real space can be determined in linear perturbation theory to be \cite{1987MNRAS.227....1K,1989ApJ...344....1G}
\begin{equation}
    P_s(k,\mu)=\left[1+\beta_d(z,k)\mu^2\right]^2P_r(k) \, ,
\end{equation}
where $\mu=\mathbf{\hat k\cdot \hat r}$ defines the cosine of the angle between the line of sight $\mathbf{r}$ and the peculiar velocity of infalling galaxies $\mathbf{k}$. The quantity $\beta_d$ is the distortion parameter, defined as
\begin{equation}
\beta_d(z,k)=\frac{f_g(z,k)}{b(z)} \, ,    
\end{equation}
where $b=\delta_g/\delta_m$ is the linear galaxy bias, which here we take to be simply described by $b(z)=\sqrt{1+z}$ \cite{Mirzatuny:2019dux}. An important feature of many modified gravity theories, specifically $f(R)$ gravity, is that the growth rate is now scale-dependent, meaning that the distortion parameter inherits the same scale-dependency, directly distinguishing it from GR. 
\begin{figure}[ht!]
    \centering
\includegraphics[width=\linewidth]{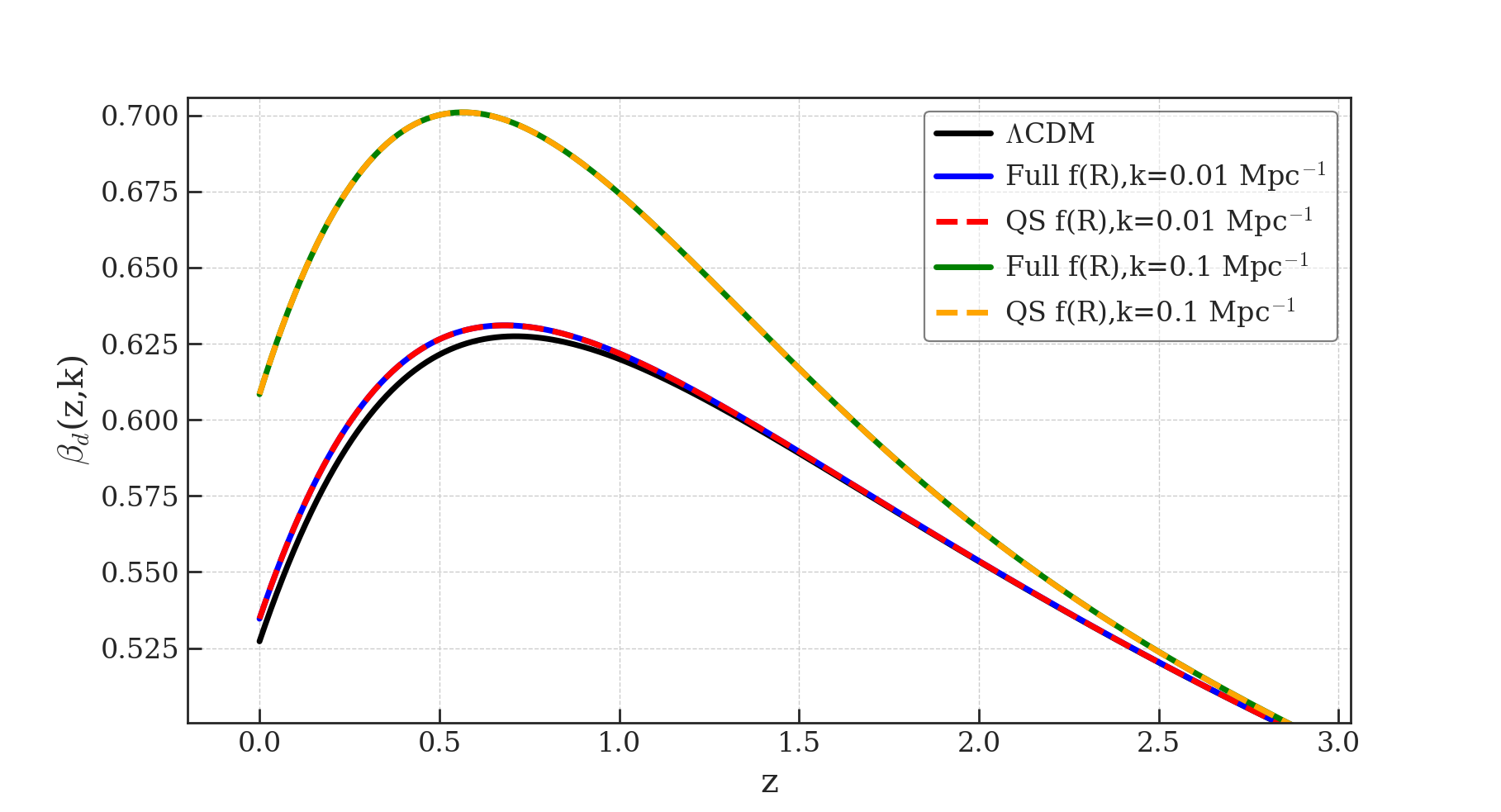}
    \caption{The distortion parameter $\beta_d(z,k)$ in GR (black) and HS $f(R)$ gravity (color). The results from the quasistatic expressions (dashed) precisely match the ones from the full expressions (full).}
    \label{fig:DistortionParameter}
\end{figure}
\par
We show the results for the distortion parameter in Fig. \ref{fig:DistortionParameter}. For all scales we see the quasistatic and full equations yield identical results, as HS mimics $\Lambda$CDM at background level. For large scales of $k\lesssim0.01\ \text{Mpc}^{-1}$ we see that even at perturbation level the deviation from $\Lambda$CDM is minimal, with the modified model being practically indistinguishable from GR until $z\sim1.0$. The difference is considerably more significant when dealing with scales $k\gtrsim0.01\ \text{Mpc}^{-1}$, with deviations from $\Lambda$CDM becoming as large as $20\%$ for redshifts near the present. This represents the biggest impact of $f(R)$ gravity at the perturbative level, as the growth rate has a sharper peak due to the higher-order nature of the theory and its consequent dependence on higher powers of $k$.

\subsection{Velocity correlation function}
The linearized conservation Eq. (\ref{eq:EtaConservationEq}) allows us to relate the components of the peculiar velocity ($\mathbf{v}=\mathbf{\nabla}v=i\mathbf{k}v$ in Fourier space) and the relative density fluctuations in the sub-Hubble limit as
\begin{equation}
    \mathbf{v}(\mathbf{k})=-i\frac{H_0}{k}f_g(z,k)\delta_m(z,\mathbf{k})\mathbf{\hat k} \, ,
\end{equation}
which is identical to GR due to the minimal coupling of $f(R)$ theories. These peculiar velocities can be studied in terms of their correlation, described by the velocity correlation tensor \cite{1988ApJ...332L...7G,1989ApJ...344....1G}
\begin{equation}
\begin{aligned}
    \Psi_{lm}(r)&\equiv\langle v_l(\mathbf{x})v_m(\mathbf{x}+\mathbf{r})\rangle\\
    &=\Psi_\perp(r)\delta_{lm}+\left[\Psi_\parallel(r)-\Psi_\perp(r)\right]\hat r_l\hat r_m \, ,
\end{aligned}
\end{equation}
where $\mathbf{r}$ denotes the separation between two points $\mathbf{r}_l$ and $\mathbf{r}_m$. In the final equality we have considered the case of a statistically homogeneous and isotropic peculiar velocity field, for which the correlation tensor can be written in terms of its parallel ($\Psi_\parallel$) and transverse ($\Psi_\perp$) components. These components can be explicitly calculated in terms of background and perturbative cosmological quantities as
\begin{align}
    \Psi_\perp(r,z)&=\frac{H_0^2}{2\pi^2}\int {\rm d}k \ f_g^2(z,k)P(z,k)\frac{j_1(kr)}{kr}\,,\\
    \Psi_\parallel(r,z)&=\frac{H_0^2}{2\pi^2}\int {\rm d}k \ f_g^2(z,k)P(z,k)\left[j_0(kr)-2\frac{j_1(kr)}{kr}\right] \,,
\end{align}
for a given matter power spectrum $P(z,k)$ and where $j_{0,1}$ are the corresponding spherical Bessel functions
\begin{equation}
    \begin{aligned}
    j_0(x)&=\frac{\sin{x}}{x}\\
        j_1(x)&=\frac{\sin{x}}{x^2}-\frac{\cos{x}}{x} \,.
    \end{aligned}
\end{equation}
Not only does the modified growth rate impact these components through its different evolution with redshift, but it leaves direct imprints of the scale-dependency of $f(R)$ gravity via its presence within the integration over $k$. The presence of the spherical Bessel functions with $kr$ as the argument transforms scales from functions wavenumber $k$ to functions of scale $r$. We thus expect that for large values of $r$ (small $k$) the modified theory's effects should be effectively negligible, such that we recover $\Lambda$CDM.

\begin{figure}[ht!]
    \centering
    \includegraphics[width=\linewidth]{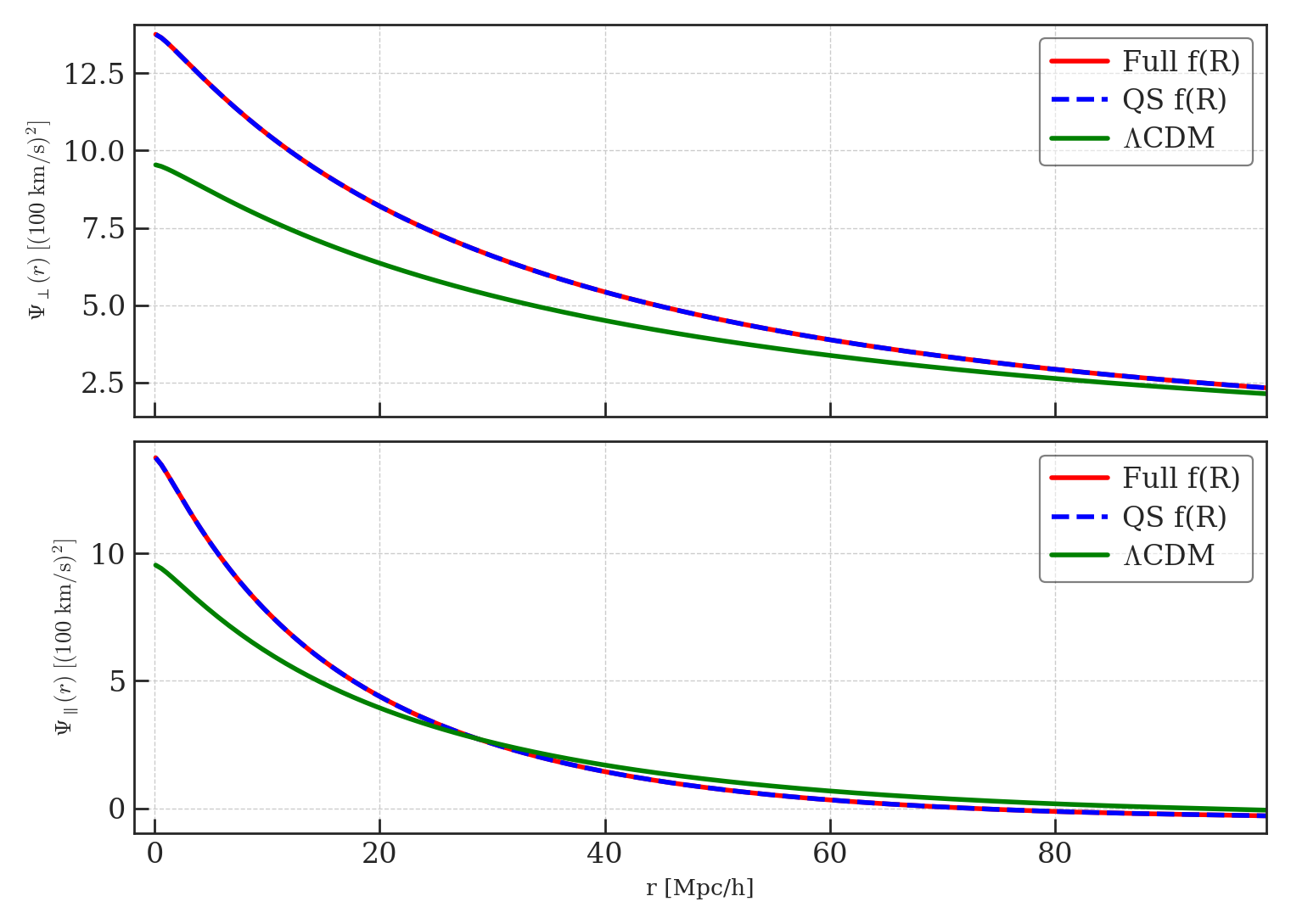}
    \caption{The transverse (top) and parallel (bottom) velocity correlation tensor components at $z=0$ in the HS $f(R)$ model and $\Lambda$CDM. The full and quasistatic $f(R)$ equations yield the same results, while both deviate from the $\Lambda$CDM prediction for small correlation distances.}
    \label{fig:CorrelationTensor}
\end{figure}
\par
We show the results for the transverse and parallel components of the correlation tensor in Fig. \ref{fig:CorrelationTensor}. We see that indeed the $f(R)$ predictions tend to the standard $\Lambda$CDM values for large distances ($r\gtrsim100\ \text{Mpc}$), as expected. As these two quantities are calculated from an integral over $k$, it is possible that for small distances the difference between the quasistatic equations and the even higher-order full equations is fleshed out in a way that allows for a clear distinction between the two, even though this has not been observed in the HS model in this work's analyzes thus far. However, we still find that the two approaches - quasistatic and full resolution - yield identical results for all correlation distances. Nevertheless, the distinct departure from $\Lambda$CDM at small scales could serve as a smoking gun for modified gravity in peculiar velocity catalogs such as SFI++ \cite{Springob:2007vb} or 6dF \cite{Springob:2014qja}, as well as providing additional means to constrain the theory's parameters.   

\subsection{Integrated Sachs-Wolfe effect}
The late-time integrated Sachs-Wolfe (ISW) effect follows from the variation over time of the scalar metric perturbations. It is directly observed via the resulting fluctuations in the CMB temperature via
\begin{equation}
    T^{\rm ISW}(\mathbf{\hat n})=2\int_0^{z_{\rm rec}}{\rm d}z \ \frac{\partial\Phi_{\rm WL}(r(z),\mathbf{\hat n})}{\partial z} \,.
\end{equation}
This can be recast in terms of the multipole moments of the temperature fluctuation field $a_{\ell m}$, allowing us to calculate the contribution of the late-time ISW effect to the radiation power spectrum \cite{Cooray:2001ab}
\begin{equation}
    C^{\rm ISW}_\ell=\langle\left| a_{\ell m} \right|^2\rangle=\frac{2}{\pi}\int_0^\infty k^2 {\rm d}k\ P(k)I_\ell^2(k) \, ,
\end{equation}
where
\begin{equation}
I_\ell(k)=\int_0^{z_{\rm rec}}{\rm d}z\ j_\ell(kr(z))\frac{{\rm d}\Phi_{\rm WL}(z,k)}{{\rm d}z}\, ,
\end{equation}
with the comoving separation explicitly calculated in terms of redshift as 
\begin{equation}
    r(z)=\int_0^z\frac{{\rm d}z'}{H(z')}\,.
\end{equation}
Note that we have kept all equations in terms of the weak lensing potential $\Phi_{\rm WL}$ and not of the density fluctuations, as in $f(R)$ this quantity is not simply related to the density contrast field $\delta$, as shown in Section \ref{sec:FRGravity}. The ISW effect alters CMB temperature maps in a way that correlates with probes of the gravitational potential from galaxy catalogs (labelled M). This correlation can be analytically computed as \cite{Cooray:2001ab}
\begin{equation}
    C_\ell^{\rm ISW-M}=\frac{2}{\pi}\int k^2 {\rm d}k\ P(k)I_\ell^{\rm ISW}(k)I_\ell^{\rm M}(k) \, ,
\end{equation}
where 
\begin{equation}
\begin{aligned}
    I_\ell^{\rm M}&=\int_0^{r_0}{\rm d}r \ W^{\rm M}(k,r)j_\ell(kr)\\
    &=\int_0^{r_0}{\rm d}r \ b(kr)n_M(r)D(r)(k,r)j_\ell(kr) \, . 
\end{aligned}
\end{equation}
Here $W^{\rm M}(k,r)$ is the window function of the tracer of the large-scale gravitational potential, with $D(z)$ denoting the growth factor and $b(k,r)$ being the scale-dependent bias, often taken to be constant and thus factored out.
The distribution of sources can be modelled by the analytic expression
$$n_M(z)=A(z/z_0)^2\exp\left[-(z/z_0)^{3/2}\right]\,, $$
with the constant $z_0$ denoting the effective depth of the catalog and being calculated from the median redshift of the source distribution as $z_m=1.412\,z_0$, while $A$ is determined from the normalization $\int n_M(z){\rm d}z=1$ \cite{Olivares:2008bx}. One can then estimate observables such as the correlation of a galaxy catalog (M) with a CMB map ($T$)
\begin{equation}\label{eq:TMCorrelation}
    \langle T*{\rm M}\rangle=\sum_\ell\frac{2\ell+1}{4\pi}C^{\rm ISW-M}_\ell P_\ell(\cos\theta) \, ,
\end{equation}
which can be compared to observations, as described in Ref. \cite{Olivares:2008bx}.
\par
\begin{figure*}[ht!]
    \centering
    \includegraphics[width=0.5\linewidth]{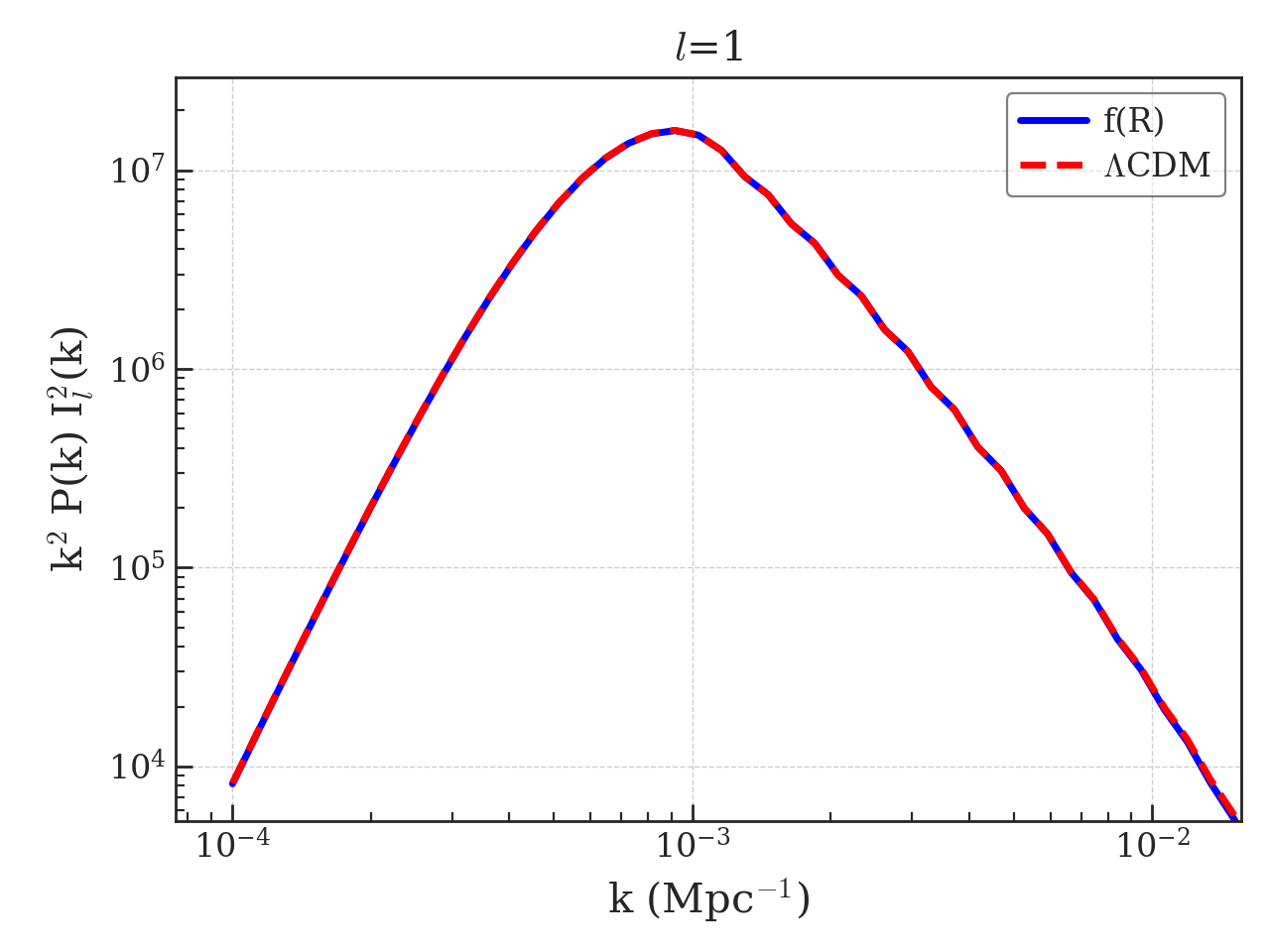}
    \includegraphics[width=0.46\linewidth]{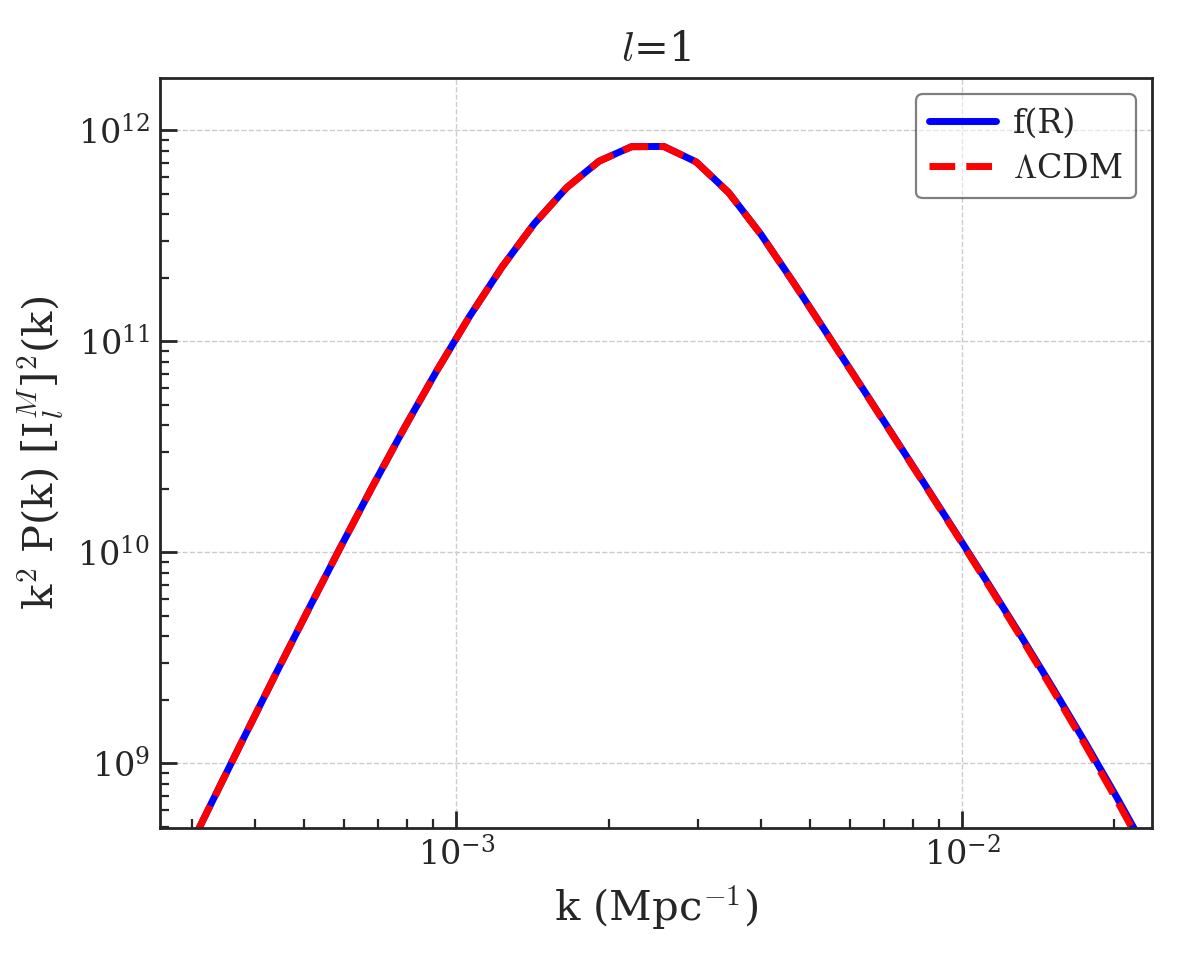}
    \caption{The integrand of the ISW (left) and ISW-M (right) power spectrum coefficients for $\ell=1$. The HS model only differs appreciably from $\Lambda$CDM beyond the each integrand's peak, rendering the modified theory's effects negligible in the calculation.}
    \label{fig:ISW}
\end{figure*}

As these quantities are calculated from a sum over various values of the multipole $\ell$, with each $C_\ell$ consuming considerable numerical resources and CPU time to determine, we start by analysing the integrand of both the ISW and ISW-M coefficients, i.e. $k^2P(k)I^{\rm ISW}_\ell(k)$ and $k^2P(k)I^{\rm M}_\ell(k)$. We show the results for $\ell=1$ in Fig. \ref{fig:ISW}. Immediately it is clear that the difference between the HS model and $\Lambda$CDM is small, with modified effects only arising at large values of $k$. However, the nature of the power spectrum and the spherical Bessel functions cause both integrands to peak around the range $k\in\left[10^{-3},10^{-2}\right]\ \text{Mpc}^{-1}$, then rapidly decreasing by various orders of magnitude beyond that range. This means that the modified theory's effects are ``washed away" in the integration and have negligible impact on the values of the $C_\ell$ coefficients and therefore on the associated observables such the one shown in Eq. (\ref{eq:TMCorrelation}). We are thus led to conclude that no conclusive evidence for or against $f(R)$ models can be drawn from observations of the late-time ISW effect.

\section{Fitting growth of structure data}\label{sec:fsigma8Fit}
As the HS $f(R)$ model is mostly unconstrained at the background level due to being able to mimic $\Lambda$CDM model's cosmic expansion history, we must rely on its perturbative effects to provide definitive constraints and evidence for or against this class of $f(R)$ modified theories. Perhaps one of the most direct measurements is that of the growth rate $f_g$, as well as the root mean squared of matter fluctuations averaged over distances of $8h^{-1}\,\text{Mpc}$, labelled $\sigma_8$ and defined as
\begin{equation}
    \label{sigma_Def}
    \sigma_8^2=\int_0^\infty k^2\frac{{\rm d}k}{2\pi^2}P(k)W^2(kR_8)\,,
\end{equation}
where $W(x)$ is the Fourier space top-hat window function
\begin{equation}
    W(x)=3\frac{\sin{x}-x\cos x}{x^2}, \,R_8=8h^{-1}\,\text{Mpc}\, .
\end{equation}
Due to the power spectrum evolving with the square of the linear growth factor as $P(z,k)\propto D^2(z)$, the $\sigma_8$ parameter evolves as $\sigma_8(z)\propto D(z)$ in $\Lambda$CDM. Of course the same is not true for general scale-dependent theories, as there $D(z,k)$ cannot be taken outside the integral over $k$ in Eq. 
\eqref{sigma_Def}. In the HS $f(R)$ model, scale-dependence is present in $D(z,k)$ although we find that it can be taken to be approximately scale-independent in the context of this integration. The same cannot be said about the growth rate $f_g(z,k)$, which has much more pronounced scale-dependence and thus provides significantly different results for theoretical predictions of $f_g\sigma_8$ depending on the considered scale. 
Due to the fact that the measurements are associated with distances of $\sim8$ Mpc, in the following we consider the reference value of $k_{\rm ref}=0.1\,\text{Mpc}^{-1}\sim(8\,\text{Mpc})^{-1}$ for the scale used in the coefficients in Eq. (\ref{eq:FullDiffEq}) \cite{Arjona:2018jhh,Arjona:2020yum}. \par
We consider 66 points from $f_g\sigma_8$ data compiled in Ref. \cite{Skara:2019usd} along with 12 points of isolated growth data data compiled in Ref. \cite{Sahlu:2024dxp}. Among the $f_g\sigma_8$ data, all points are taken to be uncorrelated as their covariance matrix is not explicitly calculated, except in the case of the WiggleZ, SDSS-IV and BOSSDR12 data, for which we use the independent covariance matrices provided in Ref. \cite{Sagredo:2018ahx}.
We also use individual $\sigma_8(z)$ measurements taken from Ref. \cite{Perenon:2019dpc}. All $\{f_g\sigma_8,\,f_g,\,\sigma_8\}$ values are shown in Tables \ref{tab:fsigma8} and \ref{tab:sigma8} in Appendix \ref{sec:DataAppendix}. Similarly to the work from Ref. \cite{Perenon:2019dpc} on Horndeski theories with the speed of gravitational waves equal to that of light, we impose that the early-time behavior of the theory should not be totally free, instead abiding by the most recent constraints on $\Omega^{\rm Planck}_{m,0}=0.315\pm0.007$ and $\sigma^{\rm Planck}_{8,0}=0.811\pm0.006$, along with their covariance matrix, from the 2018 Planck results \cite{Planck:2018vyg}. In order to compare the modified theory's prediction for $\sigma_8$ at the CMB with the Planck constraint, we must extrapolate its value from the present to the recombination epoch.
Due to the $f(R)$ modifications to $\Lambda$CDM effectively vanishing for redshifts $z\gtrsim5$, all differences will necessarily arise from this point onwards, such that we may calculate the comparison value $\sigma_{8,0}^*$ as
\begin{equation}
    \sigma_{8,0}^*=\sigma_{8,0}\frac{D_{\Lambda \text{CDM}}(z=0)}{D_{\rm HS}(z=0,k=k_{\rm ref})} \,,
\end{equation}
where $\sigma_{8,0}$ is a free parameter representing the physical value of $\sigma_8$ at present in the $f(R)$ model, which is constrained by the statistical analysis. We thus calculate the $\chi^2$ from the Planck constraints as
\begin{equation}
\begin{aligned}
    \chi^2_{\rm Planck}=\vec \Delta\cdot C^{-1}_{\rm Planck} \cdot\vec\Delta\, ,
\end{aligned}
\end{equation}
where
\begin{equation}
    \vec\Delta=\left(\Omega_{m,0}-\Omega_{m,0}^{\rm Planck},\sigma_{8,0}^*-\sigma_{8,0}^{\rm Planck}\right) \, ,
\end{equation}
thus following the same prescription as in Ref. \cite{Perenon:2019dpc}. \par
\begin{figure}[ht!]
    \centering
    \includegraphics[width=\linewidth]{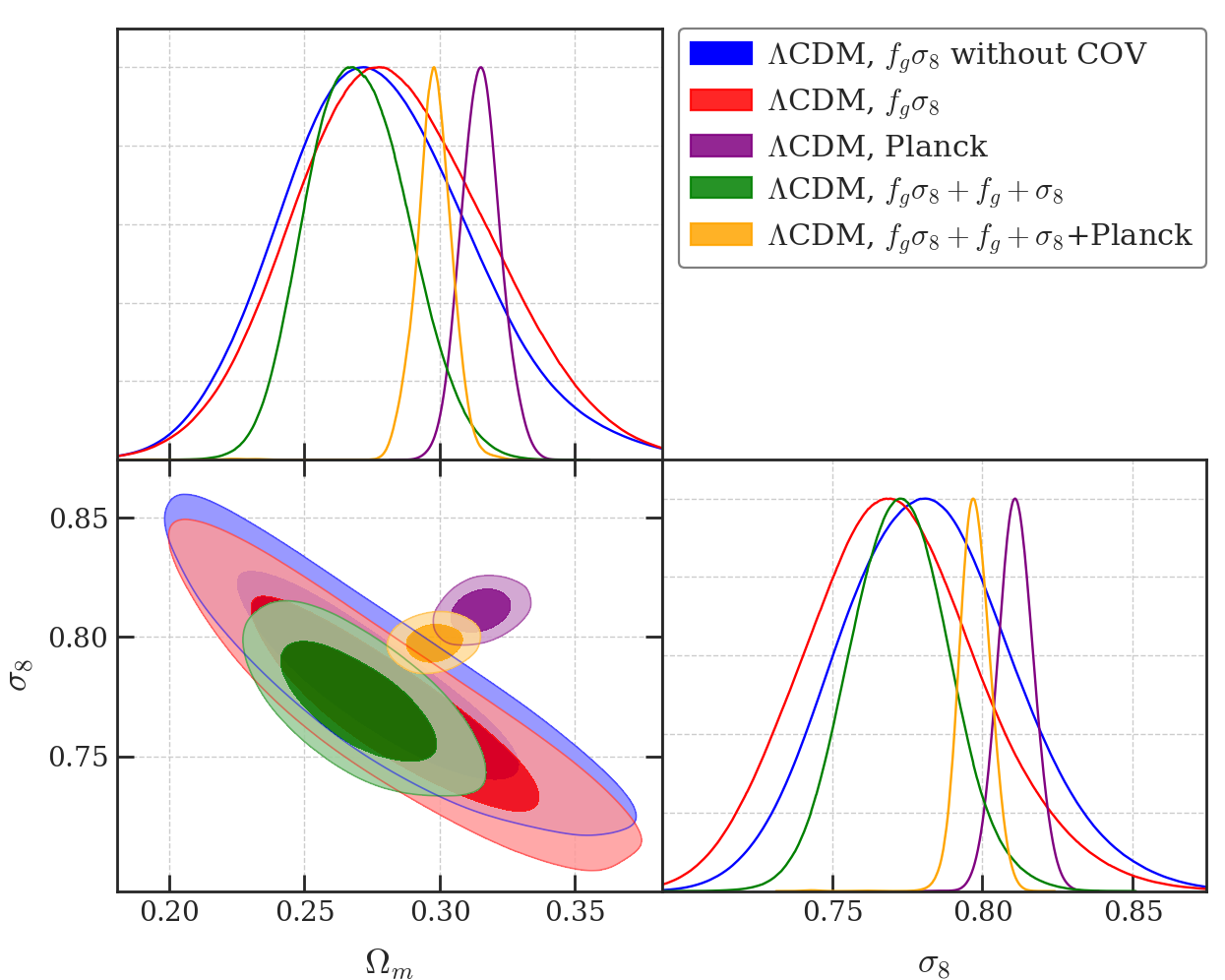}
    \caption{Posteriors for $\Lambda$CDM for $f_g\sigma_8$, both considering and disregarding the correlation between data points, a combination of $f_g\sigma_8+f_g+\sigma_8$ data, and with the addition of constraints on the parameters from Planck.}
    \label{fig:LCDM_OmegaM_sigma8_Posteriors}
\end{figure}

We used the same MCMC sampler as in Sec. \ref{subsec:BackgroundData} to determine constraints on the modified HS theory, sticking to wide uniform priors for $\Omega_m\in[0.1,0.5]$ and $\sigma_8\in[0.6,1.0]$. In the case of $\dHS$ we used a log-uniform prior over the range $\left[1,10^6\right]$ in order to cover the possibility of both strong and negligible modifications to GR. As a baseline we start by considering spatially flat $\Lambda$CDM with $\Omega_{m,0}$ and $\sigma_{8,0}$ as free parameters. The cosmological constant is estimated by $\Omega_\Lambda\approx1-\Omega_{m,0}$, where we neglect the contribution of radiation for late times. We consider both isolated $f_g\sigma_8$ data along with its combination with $f_g$ and $\sigma_8$ data. Furthermore, we analyze the effect of the presence of the covariance matrix to account for the correlation in some of the measurements, as well as the constraints from the Planck collaboration \cite{Planck:2018vyg}. These results are shown in Fig. \ref{fig:LCDM_OmegaM_sigma8_Posteriors}. It becomes clear that the inclusion of individual $f_g$ and $\sigma_8$ data considerably helps in constraining the cosmological parameters, as the posteriors visibly shrink. We also find that the inclusion of the adequate covariance matrices for correlated points has a non-negligible impact on the parameters, such that its usage is vital in obtaining the most accurate - as well as correct - results, as we do for the remainder of this section. The most significant change follows from the stringent constraints from the Planck data, which pull $\Omega_{m,0}$ from $\sim0.260$ to higher values around $\sim0.300$, as well as displacing the best fit of $\sigma_8$ to higher values, all while significantly reducing the posterior area. It should be noted that the Planck constraints are in clear tension with the $f_g\sigma_8(+f_g+\sigma_8)$ data, as the posteriors are incompatible over the $2\sigma$ level (95\% CL) when considering these datasets independently. Of course, incompatibility of early- and late-time data for cosmological parameters is nothing new, as the so-called Hubble and $\sigma_8$ tensions have been part of intense discussion in the cosmological research community for the past few years \cite{DiValentino:2021izs,Joseph:2022jsf}. This means that combining these sources of data must be done with this tension in mind, such that only conclusions drawn directly from the late-time structure growth data can be truly considered as consistent, as the Planck constraints lead to direct results for both $\Omega_m$ and $\sigma_8$ by definition. Regardless of all of these considerations, our results point to a best-fit of $\Lambda$CDM to the complete $f_g\sigma_8+f_g+\sigma_8+\text{Planck}$ dataset of $\Omega_{m,0}=0.298$ and $\sigma_{8,0}=0.795$. \par

\begin{figure*}[ht!]
    \centering
    \includegraphics[width=0.465\linewidth]{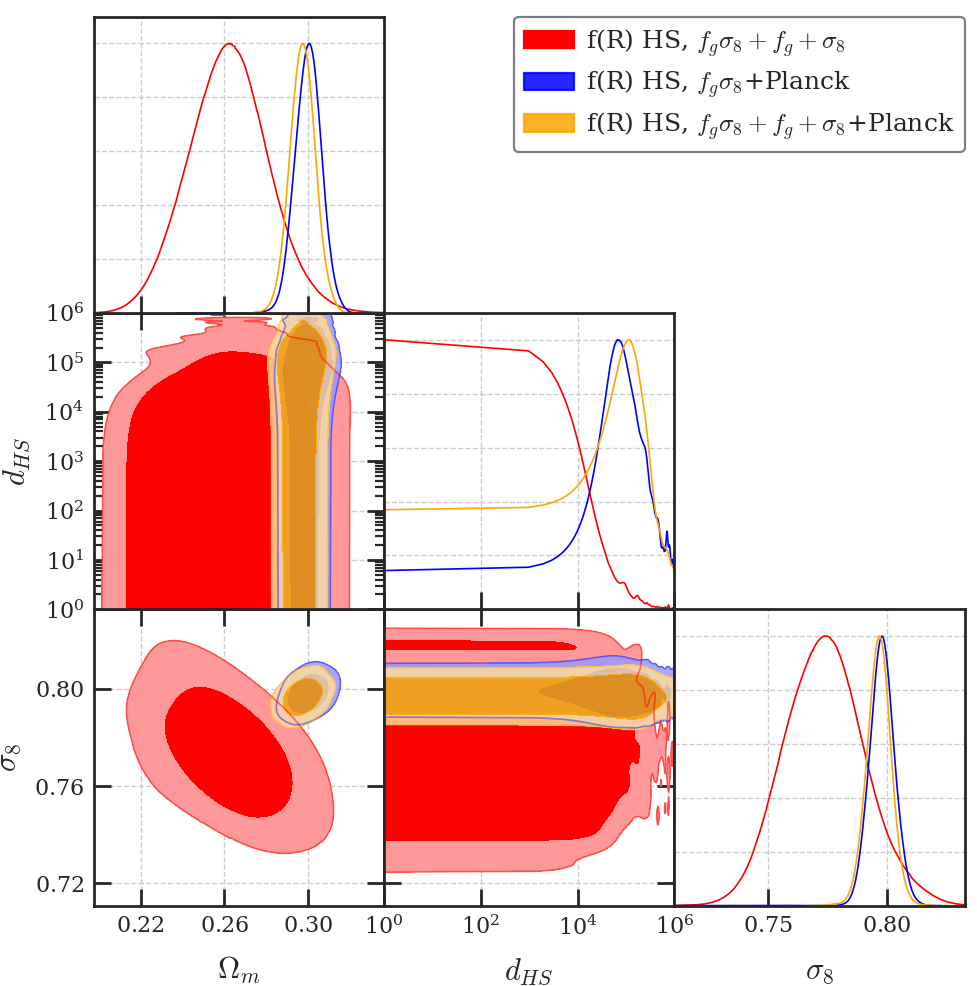}
    \includegraphics[width=0.525\linewidth]{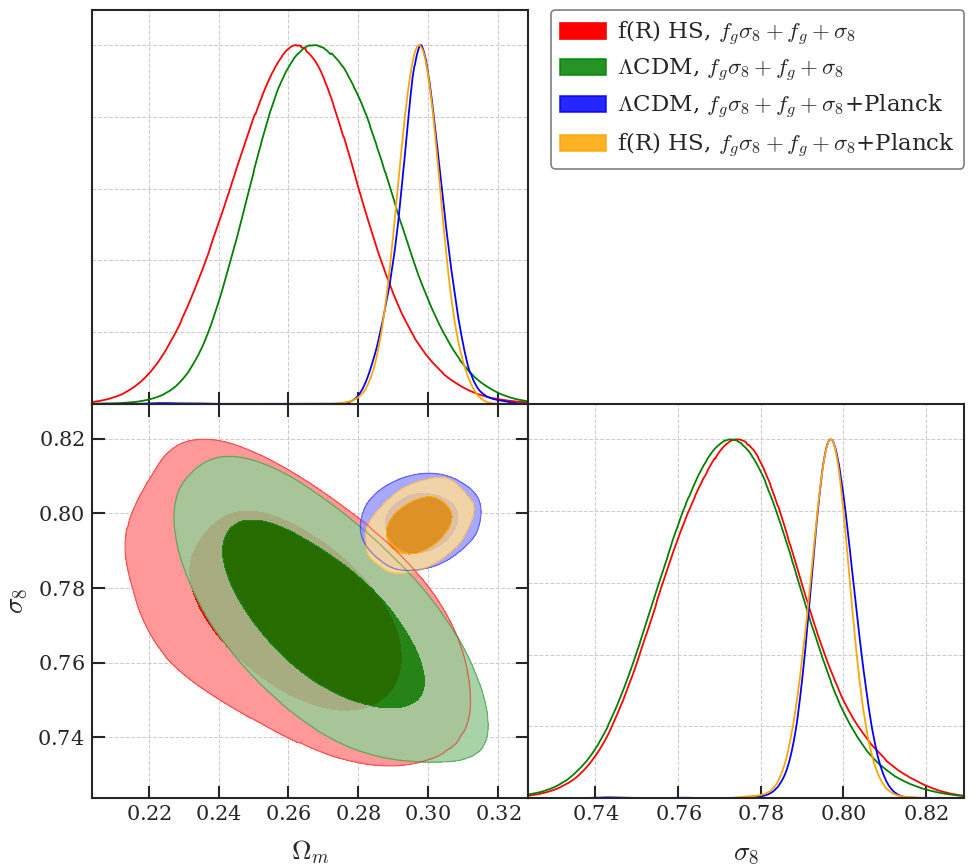}
    \caption{Left panel: Hu-Sawicki $f(R)$ model posteriors for $f_g\sigma_8$ and $f_g\sigma_8+f_g+\sigma_8$ data catalogs, both with and without Planck constraints on $\Omega_m$ and $\sigma_8$. Right panel: $\Omega_m-\sigma_8$ posteriors for both $\Lambda$CDM and the HS $f(R)$ model.}
    \label{fig:FR_OmegaM_sigma8_Posteriors}
\end{figure*}

We now analyze the HS model for both $f_g\sigma_8(+\,\text{Planck})$ and $f_g\sigma_8+f_g+\sigma_8(+\,\text{Planck})$ datasets, this time always considering the correct covariance matrix between data points, as we have already shown its importance in the discussion on $\Lambda$CDM above. We show the results for the three constrained parameters on the left panel of Fig. \ref{fig:FR_OmegaM_sigma8_Posteriors}. We can immediately see that the $\dHS$ parameter is not very accurately constrained by the MCMC chains, as found in past works on the HS model such as Ref. \cite{delaCruz-Dombriz:2015tye}, for example. The posteriors for the $f_g\sigma_8+f_g+\sigma_8$ dataset seem to indicate that lower values of $\dHS$ are favoured by the data, although the marginalized posterior ranges over several orders of magnitude in the logarithmically scaled axis. However, the most likely values are in fact higher values, as seen by the $f(R)$ and $\Lambda$CDM posteriors being practically identical for $\Omega_m-\sigma_8$ in the right plot of Fig. \ref{fig:FR_OmegaM_sigma8_Posteriors}.
Focusing instead on the datasets combined with Planck constraints, we see the posteriors indicate more clearly that stronger modifications of GR are disfavoured, as they introduce sharp changes in the evolution of the growth rate and thus make late-time and early-time data even more incompatible than in $\Lambda$CDM, such that the MCMC chains converge on weaker modifications with almost identical posteriors to the ones in Fig. \ref{fig:LCDM_OmegaM_sigma8_Posteriors}. In the right plot of Fig. \ref{fig:FR_OmegaM_sigma8_Posteriors} we see that the addition of Planck constraints leads to $f(R)$ having even less impact on the posteriors of the physical cosmological parameters $\Omega_m$ and $\sigma_8$. \par
Our results point to a preference of $\Lambda$CDM over the HS $f(R)$ model such that no evident benefit follows from considering an $f(R)$ modified action in the context of large-scale structure data. Indeed, the preferred values of $\dHS\gtrsim 10^4-10^{5}$ point to $|f_{R_0}|\lesssim 10^{-6}-10^{-5}$, compatible with other constraints already found on the HS model using samples of X-ray selected clusters combined with CMB lensing potential data \cite{Cataneo:2014kaa} and a data compilation of cosmic chronometers, baryon acoustic oscillations (BAO) and SNIa distance moduli \cite{Nunes:2016drj}. This also means that there are no significant modifications to the statistical determination of the cosmological parameters associated with late-time growth of structures, namely ($\Omega_m$,$\sigma_8$), with no visible improvement on the tension between early and late-time measurements of the $\sigma_8$ parameter \cite{Joseph:2022jsf}. We thus conclude that the HS model is severely constrained at a level that enforces it to have practically no effect on cosmological effects apart from the introduction of the cosmological constant, which is introduced \emph{a~priori} in the model, as shown in Eq. \eqref{eq:HS_Limit}.

\section{Conclusions}\label{sec:Conclusions}
In this communication we analyzed the effect of $f(R)$ metric gravity on the growth of large-scale structures in the late-time Universe. More specifically, we considered the paradigmatic $n=2$ Hu-Sawicki model, which minimally modifies the  Einsteinian action by the introduction of an $R$-dependent function that brings with it the inclusion of an effective cosmological constant driving the observed accelerated expansion of the Universe. By determining the theoretical properties of this model by resorting to the growth of large-scale structures, we predicted its cosmological implications and discussed what current data can tell us about its validity and what constraints data impose on its parameters. \par

We started by reviewing the linearized field equations in General Relativity and their implications on the density and metric perturbations to the standard $\Lambda$CDM model, including the distinction between the equations in the sub-Hubble regime and those without any assumed approximations. After introducing the framework of $f(R)$ gravity, we constrained the theory's parameter space with SNIa distance moduli from the DES collaboration, concluding that in this realm the two Hu-Sawicki parameters turn out to be correlated through the data's preference for the presence of a cosmological constant and that one may treat the 
resulting cosmic expansion as identical to $\Lambda$CDM up to a very good approximation. Furthermore, we detailed how this type of modified theory alters the general background and linearized field equations. In the latter, by departing from the correct full differential equation of density perturbations without the assumption of the quasistatic approximation or the sub-Hubble limit, we extended these results to the metric potentials, which may be calculated explicitly at each scale and redshift in terms of the relative density contrast and its time derivatives of up to third order. Using these full equations, we looked at typical reference scales of $k_{\rm ref}=0.1 \ \text{Mpc}^{-1}$ and moderately strong modifications of General Relativity and determined that the approximate $\Lambda$CDM background implies the acceptable quality of the quasistatic approximation for the obtained metric potential equations,  analogously to what was found in Ref. \cite{delaCruz-Dombriz:2008ium} for density perturbations.
\par

We then looked at the implications of these modified theories on the growth of cosmic  structures. The altered growth rate leads to a distinct peak in the redshift-space distortion parameter $\beta_d$, which is stronger for smaller scales and recovers the $\Lambda$CDM result for larger scales, as $f(R)$ gravity is scale-dependent. The choice of a particular reference scale is no longer problematic in what concerns the velocity correlation tensor, in particular its parallel and transverse components, which are calculated from an integration over all scales and thus incorporate the scale-dependence of $f(R)$ gravity into one total effect. As expected, the predictions of the modified theory match with those of $\Lambda$CDM for large correlation distances, while at smaller distances the larger growth rate leads to increased correlation tensor components, which could possibly be used to further discriminate between GR and modified gravity. We concluded this iteration of cosmological implications with the integrated Sachs-Wolfe effect, which has many similarities to the calculation of the correlation tensor, in the sense that it follows from an integration over $k$. However, the presence of other terms in the integrand that decay rapidly for small scales, on which the modified theory leads to more pronounced effects, makes it so that the late-time integrated Sachs-Wolfe effect has a negligible contribution from $f(R)$ gravity.
\par

Finally, we used the most recent growth of structure data, composed of $f_g\sigma_8$, $f$ and $\sigma_8$ measurements and spanning over 80 points, to constrain the Hu-Sawicki model ($n=2$) and to find its implications on cosmological parameters at perturbative level, as the model has no effect at background level and is thus unable to be effectively constrained with SNIa and BAO data \cite{delaCruz-Dombriz:2015tye}. We also considered constraints from the 2018 Planck collaboration measurements \cite{Planck:2018vyg}, which we found to be in considerable tension with the late-time growth data, as pointed out in other recent works \cite{Joseph:2022jsf}. All datasets pointed to stringent constraints on the emergence of notable modifications from the Hu-Sawicki model at the perturbative level, with the posteriors on the parameters $\{\Omega_m,\sigma_8\}$ being practically identical within error to those from $\Lambda$CDM and the constraints on the parameter $\dHS$ being in line with past investigations of the same model \cite{Cataneo:2014kaa,Nunes:2016drj}. From the tight constraint on $\dHS\gtrsim10^5$ (or equivalently $|f_{R_0}|\lesssim10^{-6}$), it seems clear that the Hu-Sawicki $f(R)$ theory of gravity has little to no contributions to the cosmological landscape apart from the trivial introduction of an effective cosmological constant, which is done practically by definition in its formulation.
\par

The methodology introduced in Ref. \cite{delaCruz-Dombriz:2008ium} and now extended in this work allows us to consider any type of $f(R)$ theory of gravity and accurately calculate its effects on the evolution of the perturbative (scalar) sector in cosmological scenarios. Naturally, considering more intricate models that introduce more non-trivial effects than the minimal addition of the Hu-Sawicki term to the standard gravitational action would help fully flesh out the higher-order effects present in the full perturbation Eqs. \eqref{eq:FullDiffEq}, \eqref{eq:PhiSum} and \eqref{eq:PsiSum} and thus distinguish the quasistatic approximation from the correct mathematical expressions for describing general modifications to the standard Einstein-Hilbert action. The analysis of such models is an interesting avenue of research and is thus left as the topic of future works.

\section*{Acknowledgments}
\begin{sloppypar}
The work of \textbf{M.B.V.} is supported by FCT (Fundação para a Ciência e Tecnologia, Portugal) through the grant 2024.00457.BD.
\textbf{A.~d.l.C.-D.} acknowledges support from BG20/00236 action (MCINU, Spain), NRF Grant CSUR23042798041, CSIC Grant COOPB23096, Project SA097P24 funded by Junta de Castilla y Le\'on (Spain) and Grant PID2024-158938NB-I00 funded by MCIN/AEI/\allowbreak 10.13039/\allowbreak 501100011033 and by \textit{ERDF A way of making Europe}.
\end{sloppypar}

\appendix

\section{Coefficients in metric potential equations} \label{sec:CoefficientsAppendix}
In this appendix we present the coefficients for Eqs. (\ref{eq:PhiSum}) and (\ref{eq:PsiSum}). To simplify their visualisation, here we write them in the weak $f(R)$ limit, with each coefficient being shown only at lowest order in $\FF_n$. We note that these have not been presented in the lierature so far.
All of these coefficients share a common denominator, which is given by 
\begin{equation}\label{eq:CommonDenominator}
\begin{aligned}
        \mathcal{D}(\Hub,\epsilon)=\Hub^3&\left[4F^5\FF_1^2+24F^3(\kappa_2-2)\FF_1\epsilon^2+36(2-\kappa_2)^2\epsilon^4\right.\\
        &+\left.108(\kappa_1-1)(\kappa_2-2)^2\epsilon^6\right.\\
        &+\left.324(\kappa_1-1)(\kappa_2-2)^2\epsilon^8 \right]\,.
\end{aligned}
\end{equation}
We now present the coefficients for metric perturbation $\Phi$:
\begin{equation}
\begin{split}
    \mathcal{C}_{\Phi,0} &= \frac{\Hub^3(\kappa_1-1)}{\mathcal{D}(\Hub,\epsilon)}
    \biggl[ 8F^4\FF_1^2\epsilon^2 \\
    &\quad +36F^2(\kappa_2-2)\FF_1\epsilon^4 +36(\kappa_2-2)^2\epsilon^6 \\
    &\quad +108(\kappa_1-1)(\kappa_2-2)^2\epsilon^8 \biggr] \,,
\end{split}
\end{equation}

\begin{equation}
\begin{split}
    \mathcal{C}_{\Phi,1} &= \frac{\Hub^2}{\mathcal{D}(\Hub,\epsilon)}
    \biggl[ 2F^5\FF_1^2\epsilon^2 +6F^3(\kappa_2-2)\FF_1\epsilon^4 \\
    &\quad -18F(\kappa_1-1)(\kappa_2-2)\FF_1\epsilon^6 \\
    &\quad -108(\kappa_1-1)(\kappa_2-2)^2\epsilon^8 \biggr]\, ,
\end{split}
\end{equation}

\begin{equation}
\begin{split}
    \mathcal{C}_{\Phi,2} &= \frac{\Hub}{\mathcal{D}(\Hub,\epsilon)} 
    \biggl[ 2F^5\FF_1^2\epsilon^2 +6F^3(\kappa_2-2)\FF_1\epsilon^4 \\
    &\quad +36F(\kappa_1-1)(\kappa_2-2)\FF_1\epsilon^6 \\
    &\quad +54F^4(\kappa_1-1) \Bigl\{ \bigl( 4+\kappa_1(\kappa_2-8) \\
    &\qquad -2\kappa_2+\kappa_3 \bigr)\FF_1 - (\kappa_2-2)\FF_2 \Bigr\} \biggr]\,,
\end{split}
\end{equation}

\begin{equation}
\begin{split}
    \mathcal{C}_{\Phi,3} &= \frac{1}{\mathcal{D}(\Hub,\epsilon)} 
    \biggl[ -3F^4\FF_1^2\epsilon^4 \\
    &\quad +9F^3(1-3\kappa_1+\kappa_2)\FF_1^2\epsilon^6 \\
    &\quad -54F^2(\kappa_1-1)(\kappa_2-2)\FF_1\epsilon^8 \biggr] \, .
\end{split}
\end{equation}
The coefficients for the metric perturbation $\Psi$ are
\begin{equation}
\begin{split}
    \mathcal{C}_{\Psi,0} &= \frac{\Hub^3(\kappa_1-1)}{\mathcal{D}(\Hub,\epsilon)}
    \biggl[ 12F^2(\kappa_2-2)\FF_1\epsilon^4 \\
    &\quad +36(\kappa_2-2)^2\epsilon^6 \\
    &\quad +108(\kappa_1-1)(\kappa_2-2)^2\epsilon^8 \biggr] \,,
\end{split}
\end{equation}

\begin{equation}
\begin{split}
    \mathcal{C}_{\Psi,1} &= \frac{\Hub^2}{\mathcal{D}(\Hub,\epsilon)}
    \biggl[ -2F^5\FF_1^2\epsilon^2 -6F^3(\kappa_2-2)\FF_1\epsilon^4 \\
    &\quad -90F(\kappa_1-1)(\kappa_2-2)\FF_1\epsilon^6 \\
    &\quad -108(\kappa_1-1)(\kappa_2-2)^2\epsilon^8 \biggr]\, ,
\end{split}
\end{equation}

\begin{equation}
\begin{split}
    \mathcal{C}_{\Psi,2} &= \frac{\Hub}{\mathcal{D}(\Hub,\epsilon)}
    \biggl[ -2F^5\FF_1^2\epsilon^2 -6F^3(\kappa_2-2)\FF_1\epsilon^4 \\
    &\quad +9F^2\bigl( 12 -12\kappa_1 +\kappa_1^2 +3\kappa_2 -\kappa_3 \bigr)\FF_1^2\epsilon^6 \\
    &\quad +54F^4(\kappa_1-1) \Bigl\{ \bigl( 8+\kappa_1(\kappa_2-8) \\
    &\qquad -4\kappa_2+\kappa_3 \bigr)\FF_1 -(\kappa_2-2)\FF_2 \Bigr\} \biggr]\,,
\end{split}
\end{equation}

\begin{equation}
\begin{split}
    \mathcal{C}_{\Psi,3} &= \frac{1}{\mathcal{D}(\Hub,\epsilon)}
    \biggl[ 3F^4\FF_1^2\epsilon^4 \\
    &\quad +9F^3(1-3\kappa_1+\kappa_2)\FF_1^2\epsilon^6 \\
    &\quad -54F^2(\kappa_1-1)(\kappa_2-2)\FF_1\epsilon^8 \biggr] \, .
\end{split}
\end{equation}

\section{Data used in the analysis}
\label{sec:DataAppendix}

In this Appendix, we present the data used in the analysis of the Hu-Sawicki model against $f_g\sigma_8$ (combined with distinct $f_g$ and $\sigma_8$ datasets) shown in Section \ref{sec:fsigma8Fit}.

\begin{table*}[t!]
    \centering
    \caption{$f_g\sigma_8$ measurements from Ref.\,\cite{Skara:2019usd}.}
    \label{tab:fsigma8}
    \setlength{\tabcolsep}{0pt} 
    \renewcommand{\arraystretch}{1.2} 
    \begin{tabular*}{\textwidth}{@{\extracolsep{\fill}} cc cc cc cc}
        \toprule
        $z$ & $f_g\sigma_8$ & $z$ & $f_g\sigma_8$ & $z$ & $f_g\sigma_8$ & $z$ & $f_g\sigma_8$ \\
        \midrule
        0.35 & $0.440\pm0.050$   & 0.77  & $0.490\pm0.180$  & 0.17  & $0.510\pm0.060$   & 0.02  & $0.314\pm0.048$   \\
        0.02 & $0.398\pm0.065$   & 0.25  & $0.3512\pm0.0583$& 0.37  & $0.4602\pm0.0378$ & 0.25  & $0.3665\pm0.0601$ \\
        0.37 & $0.4031\pm0.0586$ & 0.44  & $0.413\pm0.080$  & 0.60  & $0.390\pm0.063$   & 0.73  & $0.437\pm0.072$   \\
        0.067& $0.423\pm0.055$   & 0.30  & $0.407\pm0.055$  & 0.40  & $0.419\pm0.041$   & 0.50  & $0.427\pm0.043$   \\
        0.60 & $0.433\pm0.067$   & 0.80  & $0.470\pm0.080$  & 0.35  & $0.429\pm0.089$   & 0.18  & $0.360\pm0.090$   \\
        0.38 & $0.440\pm0.060$   & 0.32  & $0.384\pm0.095$  & 0.32  & $0.480\pm0.100$   & 0.57  & $0.417\pm0.045$   \\
        0.15 & $0.490\pm0.145$   & 0.10  & $0.370\pm0.130$  & 1.40  & $0.482\pm0.116$   & 0.59  & $0.488\pm0.060$   \\
        0.38 & $0.497\pm0.045$   & 0.51  & $0.458\pm0.038$  & 0.61  & $0.436\pm0.034$   & 0.38  & $0.477\pm0.051$   \\
        0.51 & $0.453\pm0.050$   & 0.61  & $0.410\pm0.044$  & 0.76  & $0.440\pm0.040$   & 1.05  & $0.280\pm0.080$   \\
        0.32 & $0.427\pm0.056$   & 0.57  & $0.426\pm0.029$  & 0.727 & $0.296\pm0.0765$ & 0.02  & $0.428\pm0.0465$ \\
        0.60 & $0.480\pm0.120$   & 0.86  & $0.480\pm0.100$  & 0.60  & $0.550\pm0.120$   & 0.86  & $0.400\pm0.110$   \\
        0.10 & $0.480\pm0.160$   & 0.001 & $0.505\pm0.085$  & 0.85  & $0.450\pm0.110$   & 0.31  & $0.469\pm0.098$   \\
        0.36 & $0.474\pm0.097$   & 0.40  & $0.473\pm0.086$  & 0.44  & $0.481\pm0.076$   & 0.48  & $0.482\pm0.067$   \\
        0.52 & $0.488\pm0.065$   & 0.56  & $0.482\pm0.067$  & 0.59  & $0.481\pm0.066$   & 0.64  & $0.486\pm0.070$   \\
        0.10 & $0.376\pm0.038$   & 1.52  & $0.420\pm0.076$  & 1.52  & $0.396\pm0.079$   & 0.978 & $0.379\pm0.176$   \\
        1.23 & $0.385\pm0.099$   & 1.526 & $0.342\pm0.070$  & 1.944 & $0.364\pm0.106$   &       &                  \\
        \bottomrule
    \end{tabular*}
\end{table*}

\begin{table*}[t!]
    \centering
    \caption{$f(z)$ measurements from Ref.\,\cite{Sahlu:2024dxp} (left) and $\sigma_8$ measurements from Refs.\,\cite{Pezzotta:2016gbo,delaTorre:2016rxm,Shi:2017qpr} (right).}
    \label{tab:combined_f_sigma}
    
    \begin{minipage}[t]{0.48\textwidth}
        \centering
        \renewcommand{\arraystretch}{1.2}
        \begin{tabular}{cc}
            \toprule
            $z$    & $f_g(z)$ \\
            \midrule
            0.013 & $0.56 \pm 0.07$  \\
            0.10  & $0.464 \pm 0.04$ \\
            0.15  & $0.490 \pm 0.145$ \\
            0.18  & $0.49  \pm 0.12$ \\
            0.22  & $0.6   \pm 0.10$ \\
            0.35  & $0.7   \pm 0.18$ \\
            0.41  & $0.7   \pm 0.07$ \\
            0.55  & $0.75  \pm 0.18$ \\
            0.60  & $0.73  \pm 0.07$ \\
            0.60  & $0.93  \pm 0.22$ \\
            0.77  & $0.91  \pm 0.36$ \\
            1.40  & $0.99  \pm 0.19$ \\
            \bottomrule
        \end{tabular}
        \label{tab:f} 
    \end{minipage}
    \hfill 
    \begin{minipage}[t]{0.48\textwidth}
        \centering
        \renewcommand{\arraystretch}{1.2}
        \begin{tabular}{cc}
            \toprule
            $z$   & $\sigma_8$ \\
            \midrule
            0.10 & $0.769 \pm 0.105$ \\
            0.60 & $0.52  \pm 0.06$  \\
            0.86 & $0.48  \pm 0.04$  \\
            \bottomrule
        \end{tabular}
        \label{tab:sigma8}
    \end{minipage}
\end{table*}

\clearpage
\bibliographystyle{elsarticle-num-names} 
\bibliography{References.bib}

\end{document}